\documentclass[ALICE,manyauthors]{cernphprep}
\usepackage[comma,square,numbers,sort&compress]{natbib}
\usepackage{hyperref}
\usepackage{lineno}
\usepackage{xspace,xcolor}
\usepackage{multirow}
\usepackage{upgreek}
\usepackage[T1]{fontenc}
\usepackage{orcidlink}

\newcommand{\pp}           {pp\xspace}

\newcommand{\PbPb}         {\mbox{Pb--Pb}\xspace}

\newcommand{\pPb}          {\mbox{p--Pb}\xspace}

\newcommand{\s}            {\ensuremath{\sqrt{s}}\xspace}
\newcommand{\snn}          {\ensuremath{\sqrt{s_{\mathrm{NN}}}}\xspace}
\newcommand{\pt}           {\ensuremath{p_{\rm T}}\xspace}

\newcommand{\ycms}         {\ensuremath{y_{\rm cms}}\xspace}
\newcommand{\ylab}         {\ensuremath{y_{\rm lab}}\xspace}
\newcommand{\yfid}         {\ensuremath{y_{\rm fid}}\xspace}

\newcommand{\RpPb}         {\ensuremath{R_{\rm pPb}}\xspace}

\newcommand{\Lint}         {\ensuremath{\mathcal{L}_\mathrm{int}}\xspace}

\newcommand{\nineH}        {$\sqrt{s}~=~0.9$~Te\kern-.1emV\xspace}
\newcommand{\seven}        {$\sqrt{s}~=~7$~Te\kern-.1emV\xspace}
\newcommand{\twoH}         {$\sqrt{s}~=~0.2$~Te\kern-.1emV\xspace}
\newcommand{\twosevensix}  {$\sqrt{s}~=~2.76$~Te\kern-.1emV\xspace}
\newcommand{\five}         {$\sqrt{s}~=~5.02$~Te\kern-.1emV\xspace}
\newcommand{\thirteen}     {$\sqrt{s}~=~13$~Te\kern-.1emV\xspace}
\newcommand{\twosevensixnn}{$\sqrt{s_{\mathrm{NN}}}~=~2.76$~Te\kern-.1emV\xspace}
\newcommand{\fivenn}       {$\sqrt{s_{\mathrm{NN}}}~=~5.02$~Te\kern-.1emV\xspace}
\newcommand{\eightnn}       {$\sqrt{s_{\mathrm{NN}}}~=~8.16$~Te\kern-.1emV\xspace}

\newcommand{\GeVc}         {Ge\kern-.1emV/$c$\xspace}
\newcommand{\MeVc}         {Me\kern-.1emV/$c$\xspace}
\newcommand{\TeV}          {Te\kern-.1emV\xspace}
\newcommand{\GeV}          {Ge\kern-.1emV\xspace}
\newcommand{\MeV}          {Me\kern-.1emV\xspace}
\newcommand{\GeVmass}      {Ge\kern-.1emV/$c^2$\xspace}
\newcommand{\MeVmass}      {Me\kern-.1emV/$c^2$\xspace}

\newcommand{\ITS}          {\rm{ITS}\xspace}
\newcommand{\TOF}          {\rm{TOF}\xspace}

\newcommand{\TPC}          {\rm{TPC}\xspace}

\newcommand{\VZERO}        {\rm{V0}\xspace}

\newcommand{\ee}           {\ensuremath{\rm e^{+}e^{-}}\xspace} 
\newcommand{\pip}          {\ensuremath{\pi^{+}}\xspace}

\newcommand{\kam}          {\ensuremath{\rm{K}^{-}}\xspace}

\newcommand{\proton}    {\mbox{$\mathrm {p}$}\xspace}

\newcommand{\Lb}{\ensuremath{\rm {\Lambda_b^{0}}}\xspace}

\newcommand{\lambdac}     {\ensuremath{\mathrm {\Lambda_{c}^{+}}}\xspace}
\newcommand{\xicz}        {\ensuremath{\mathrm {\Xi_{c}^{0}}}\xspace}

\newcommand{\rmLambdas}         {\ensuremath{\mathrm {\Lambda \kern-0.2em + \kern-0.2em \overline{\Lambda}}}\xspace}

\newcommand{\Kzs}               {\ensuremath{\mathrm {K^0_S}}\xspace}

\newcommand{\Dzero}{\ensuremath{\mathrm {D^0}}\xspace}

\newcommand{\Dstar}{\ensuremath{\rm D^{*+}}\xspace}
\newcommand{\Dplus}{\ensuremath{\rm D^+}\xspace}

\newcommand{\Lcplus}{\lambdac}
\newcommand{\Lc}         {\Lcplus}
\newcommand{\LcD} {\ensuremath{\lambdac/\Dzero}\xspace}

\renewcommand{\d}{\ensuremath{\mathrm{d}}\xspace}

\begin{document}

\begin{titlepage}

\PHyear{2024}
\PHnumber{193}
\PHdate{12 July}

\title{Measurement of beauty production via non-prompt charm hadrons in p--Pb collisions at $\mathbf{\sqrt{\textit{s}_{\mathrm{NN}}} = 5.02}$ TeV}
\ShortTitle{Non-prompt charm-hadron production in p--Pb collisions}

\Collaboration{ALICE Collaboration\thanks{See Appendix~\ref{app:collab} for the list of collaboration members}}
\ShortAuthor{ALICE Collaboration}

\begin{abstract}
The production cross sections of $\mathrm {D^0}$, $\mathrm {D^+}$, and $\mathrm {\Lambda_{c}^{+}}$ hadrons originating from beauty-hadron decays (i.e.~non-prompt) were measured for the first time at midrapidity in proton--lead (p--Pb) collisions at the center-of-mass energy per nucleon pair of $\sqrt{s_{\mathrm{NN}}} = 5.02$ TeV.
Nuclear~modification~factors ($R_{\mathrm {pPb}}$) of non-prompt $\mathrm {D^0}$, $\mathrm {D^+}$, and $\mathrm {\Lambda_{c}^{+}}$ are calculated as a function of the transverse momentum ($p_{\mathrm T}$) to investigate the modification of the momentum spectra measured in p--Pb collisions with respect to those measured in proton--proton (pp) collisions at the same energy. The $R_{\mathrm {pPb}}$ measurements are compatible with unity and with the measurements in the prompt charm sector, and do not show a significant $p_{\mathrm T}$ dependence. The $p_{\mathrm T}$-integrated cross sections and $p_{\mathrm T}$-integrated $R_{\mathrm {pPb}}$ of non-prompt $\mathrm {D^0}$ and $\mathrm {D^+}$ mesons are also computed by extrapolating the visible cross sections down to $p_{\mathrm T}$ = 0. The non-prompt D-meson $R_{\mathrm {pPb}}$ integrated over $p_{\mathrm T}$ is compatible with unity and with model calculations implementing modification of the parton distribution functions of nucleons bound in nuclei with respect to free nucleons.~The non-prompt $\mathrm {\Lambda_{c}^{+}/D^{0}}$ and $\mathrm{D^+/D^0}$ production ratios are computed to investigate hadronisation mechanisms of beauty quarks into mesons and baryons. The measured ratios as a function of $p_{\mathrm T}$ display a similar trend to that measured for charm hadrons in the same collision system.

\end{abstract}
\end{titlepage}

\setcounter{page}{2}

\section{Introduction}

Measurements of heavy-flavour hadron production in hadronic collisions provide crucial tests for calculations based on quantum chromodynamics (QCD).
Due to their large masses with respect to the QCD energy scale, heavy quarks (i.e.~charm and beauty) are primarily produced at the early stages of the collision via hard-scattering processes with large momentum transfer, legitimising the calculations of inclusive production cross sections via perturbative QCD (pQCD).
These calculations rely on a factorisation scheme where the \pt-differential production cross sections of charm or beauty hadrons are calculated as a convolution of three terms: (i) the parton distribution functions (PDFs) of the incoming nucleons, which describe the Bjorken-$x$ distributions of quarks and gluons within the incoming hadrons, (ii) the partonic scattering cross section, calculated as a perturbative series in powers of the strong coupling constant $\alpha_{\mathrm{S}}$, and (iii) the fragmentation function parametrising the non-perturbative evolution of a heavy quark into a given heavy-flavour hadron species.~The fragmentation functions are determined from measurements in \ee collisions~\cite{Gladilin:2014tba} and used to compute the production cross section in hadronic collisions, under the assumption that the relevant hadronisation processes are ``universal'', i.e.~independent of the collision energy and system. 

To isolate the effects of hadronisation, heavy-flavour hadron-to-hadron production yield ratios are especially effective, since the PDFs and the partonic interaction cross sections are common to all charm or beauty hadron species and their effects cancel out in the yield ratios when using the factorisation approach. Measurements of non-strange charm and beauty-meson production cross sections in \pp and p--Pb collisions at the LHC~\cite{ALICE:2021mgk, ALICE:2017olh, LHCb:2016ikn, CMS:2021lab, ALICE:2023sgl, ALICE:2016yta, CMS:2022wkk, ATLAS:2015igt, ATLAS:2013cia, ATLAS:2015esn, ATLAS:2019jpi} show that the meson-to-meson ratios are described by the pQCD calculations at next-to-leading order accuracy with all-order resummation of next-to-leading logarithms, such as FONLL~\cite{Cacciari:1998it, Cacciari:2012ny} and GM-VFNS~\cite{Kniehl:2004fy, Kniehl:2005mk, Kniehl:2012ti, Helenius:2018uul}, and by PYTHIA 8 event generator using the Monash tune~\cite{Sjostrand:2014zea,Skands:2014pea}, which is tuned on \ee collisions.~However, all these calculations largely underpredict the production of charm and beauty baryons~\cite{ALICE:2022exq, ALICE:2021bli, ALICE:2022cop, LHCb:2023wbo}. In addition, charm and beauty baryon-to-meson yield ratios, measured at mid- and forward rapidity at the LHC, show significant deviations from the values measured in \ee collisions~\cite{ALICE:2017thy, ALICE:2020wla,  ALICE:2020wfu,CMS:2019uws,ALICE:2021psx, ALICE:2021bli, ALICE:2021rzj, ALICE:2022cop, ALICE:2022exq, LHCb:2018weo,LHCb:2019fns,LHCb:2015qvk, LHCb:2023wbo, LHCb:2019avm,ALICE:2023wbx}, indicating that the assumption of universality of the hadronisation process across collision systems might no longer be valid at the LHC~\cite{ALICE:2023sgl, ALICE:2021dhb, Cacciari:2012ny}.~The reconstruction of prompt charm hadrons, which are produced from the decay of excited charm states or from charm-quark hadronisation, and of non-prompt charm hadrons, which stem from the decay of beauty hadrons, provides a good approach for probing the distinct sectors of charm and beauty.~The prompt charm baryon-to-meson production ratios were measured in \pPb collisions by the LHCb Collaboration at both forward (\mbox{$1.5 < \ylab < 4.0$} in the laboratory-frame) and backward (\mbox{$-5.0 < \ylab < -2.5$}) rapidity regions~\cite{LHCb:2018weo,LHCb:2023cwu}. Comparatively, these findings indicate an augmented baryon-to-meson yield ratios measured at forward/backward rapidity with respect to the corresponding measurements in \ee and ep collisions, although this is smaller compared to the enhancement observed when considering midrapidity measurements~\cite{ALICE:2020wla, ALICE:2021psx, ALICE:2021bli}.~In the beauty sector, the ALICE Collaboration measured production cross sections of non-prompt \Dzero and \Lc hadrons at midrapidity ($|y| < $ 0.5) in pp collisions at $\sqrt{s}$ = 13 TeV~\cite{ALICE:2023wbx}. The measured baryon-to-meson production ratio shows an enhancement similar to that observed in the charm sector, and the enhancement at midrapidity is similar to the one observed at forward rapidity by the LHCb Collaboration measuring the \Lb-baryon production relative to that of B mesons in pp and \pPb collisions~\cite{LHCb:2019fns,LHCb:2015qvk, LHCb:2023wbo, LHCb:2019avm}. Modification of charm and beauty baryon-to-meson ratios from \ee to \pp and p--Pb collisions suggests the influence of the hadronic or partonic environment on the hadronisation process~\cite{Altmann:2024kwx}. Further hadronisation effects, apart from pure in-vacuum fragmentation, like recombination (or coalescence) of charm quarks with quarks or di-quarks from a thermal medium~\cite{Minissale:2020bif,Minissale:2024gxx,Song:2018tpv, Beraudo:2023nlq}, statistical hadronisation including contributions from undiscovered higher-mass resonant states~\cite{Andronic:2021erx,He:2019tik,He:2022tod}, and string formation beyond the leading-colour approximation~\cite{Christiansen:2015yqa,Bierlich:2023okq}, serve as examples of implementations considered by theorists to refine the modelling of hadronisation to baryons. 

Measurements of heavy-flavour hadron production in proton--nucleus collisions also allow to study various effects related to the presence of nuclei in the colliding system, denoted as cold-nuclear-matter (CNM) effects. In the initial state of the collisions, the PDFs of inbound nucleons are modified by the nuclear environment as compared to free nucleons, depending on the parton momentum fraction $x$, the squared momentum transfer $Q^2$ in the hard scattering processes, and the nucleus mass number $A$~\cite{Arneodo:1992wf, Malace:2014uea}. At LHC energies and midrapidity ($|\ylab|<0.5$), the most relevant effect on the PDF is called shadowing. It corresponds to a reduction of the parton densities at $x$ lower than $10^{-2}$, which becomes stronger when $Q^2$ decreases and the nucleus mass number $A$ increases.~This effect, induced by the high phase-space density of small-$x$ partons~\cite{Eskola:2009uj, Hirai:2007sx, deFlorian:2003qf, Eskola:2016oht}, can be described within the factorisation scheme by means of phenomenological parametrisations, denoted as nuclear PDFs (nPDFs).  
The modification of the small-$x$ parton dynamics can significantly reduce the charm and beauty hadron yield with respect to pp collisions at low \pt. Furthermore, multiple scattering of partons in the nucleus can modify the kinematic distribution of the produced hadrons.~Partons can lose energy in the initial stages of the collision via initial-state radiation~\cite{Vitev:2007ve} or experience transverse momentum broadening due to multiple soft collisions before the heavy-quark pair is produced~\cite{Wang:1998ww, Kopeliovich:2002yh}. These initial-state effects are expected to have an influence on charm-hadron production at low and intermediate $\pt$ ($\pt < 4$~\GeV/$c$). For this reason, measurements of the charm- and beauty-hadron production cross section and its nuclear modification factor \RpPb, which is defined as the ratio of the production cross section in \pPb to that in \pp collisions scaled by the mass number of the Pb nucleus ($A_{\mathrm{Pb}}$), down to low \pt could provide important information, helping to significantly reduce the uncertainties on the gluon nPDFs at small $x$~\cite{Kusina:2017gkz, Eskola:2019bgf}. 

In addition to the aforementioned initial-state effects, final-state effects may also be responsible for modifications of heavy-flavour hadron yields and momentum distributions. 
Measurements in the light- and heavy-flavour sectors in high-multiplicity pp and p--Pb collisions at different collision energies showed significant flow-like effects~\cite{ALICE:2013wgn, CMS:2013pdl,ALICE:2022exq}. These effects resemble those observed in high-energy nucleus--nucleus collisions and are ascribed to quark--gluon plasma formation.
In this picture, particles of larger mass are boosted to higher transverse momenta due to a common velocity field~\cite{PhysRevC.48.2462}. However, baryon production at intermediate \pt may also be enhanced as a result of hadronisation via quark recombination~\cite{PhysRevLett.90.202303}.

The ALICE and CMS Collaborations measured the \RpPb of D and B meson in \pPb collisions, finding values close to unity within the rapidity ranges $|\ylab| <$ 0.5~\cite{ALICE:2019fhe} and $|\ylab| <$ 2.4~\cite{CMS:2015sfx}, respectively. In contrast, the LHCb Collaboration measurements at forward (2.5 $< \ylab <$ 3.5) and backward rapidity (-3.5 $< \ylab <$ -2.5)~\cite{LHCb:2019avm}, evidence a suppression of up to 20\% for beauty mesons in the forward rapidity interval and no significant suppression in the backward rapidity interval. Model calculations based on nPDFs describe well these observations.

In the baryon sector, the ALICE Collaboration found that the \Lc \RpPb depends on \pt, being below unity at low \pt and above unity at high \pt~\cite{ALICE:2020wla}. Simulations based on POWHEG+PYTHIA 6~\cite{Frixione:2007nw,Sjostrand:2006za}, combined with EPPS16 nPDF~\cite{Eskola:2016oht}, reproduce the results at low \pt but do not describe the measured trend at intermediate \pt. The \Lb measurements in \pPb collisions at large rapidities by the LHCb Collaboration are consistent with the corresponding measurements in pp collisions within uncertainties~\cite{LHCb:2019avm}. 

In this article, possible effects related to the modification of hadronisation mechanisms, and initial and final-state effects at midrapidity ($|y_{\rm lab}| < 0.5$) in \pPb collisions in the beauty sector are investigated. The \pt-differential production cross sections and nuclear modification factors of non-prompt \Dzero, \Dplus, and \Lc hadrons in \pPb collisions at $\snn = 5.02$ TeV are reported. The \Dzero meson is reconstructed in the interval 1 $< \pt <$ 24 \GeVc, while the \Dplus and \Lc hadrons are reconstructed in the interval \mbox{2 $< \pt <$ 24 \GeVc}. By integrating the \pt-differential results and extrapolating to \pt = 0 using pQCD calculations, the \pt-integrated non-prompt \Dzero and \Dplus production cross sections are computed. The paper is organised as follows. Section~\ref{sec:experiment_data_sample} describes the ALICE apparatus and the analysed data samples. Section~\ref{analysisTech} details the analysis methods used and outlines the corrections applied to calculate the \pt-differential production cross sections. Section~\ref{systUnc} describes the sources of systematic uncertainty. The results are presented in Section~\ref{result}. Finally, a summary is given in Section~\ref{summary}.

\section{Experimental setup and data sample}
\label{sec:experiment_data_sample}

The ALICE apparatus~\cite{ALICE:2008ngc} consists of a set of detectors for particle reconstruction and identification at midrapidity ($| \eta| < 0.9$) embedded in a solenoidal magnet, a forward ($-4 < \eta < -2.5$) muon spectrometer, and a set of forward and backward detectors for triggering and event characterisation. Typical detector performance in \pp, \pPb, and \PbPb collisions is presented in~\cite{ALICE:2014sbx}.~The reconstruction of heavy-flavour hadrons from their hadronic decay products at midrapidity primarily relies on the Inner Tracking System (ITS)~\cite{ALICE:2013nwm}, the Time Projection Chamber (TPC)~\cite{Alme_2010}, and the Time-Of-Flight detector (TOF)~\cite{Akindinov:2013tea} for tracking, primary and decay vertex reconstruction, and charged-particle identification (PID). The V0 detector arrays~\cite{ALICE:2013axi} are used for triggering and event selection.

The data sample used in this analysis are from proton--lead collisions at $\snn = 5.02$ TeV collected in 2016. The events were recorded with a minimum-bias (MB) interaction trigger that required coincident signals in both scintillator arrays of the \VZERO detector, which covers the full azimuth in the pseudorapidity intervals $-3.7 < \eta < -1.7$ and $2.8 < \eta < 5.1$. The \VZERO timing information was used together with that from the Zero-Degree Calorimeter (ZDC)~\cite{ALICE:2014sbx} for offline rejection of beam-beam or beam-gas interactions occurring outside the nominal colliding bunches. 

To ensure uniform acceptance in pseudorapidity, events were required to have a reconstructed collision vertex located within $\pm 10$ cm from the nominal collision point along the beam-line direction.~Events composed of several interactions per bunch crossing, whose probability was below 0.5\%, were rejected using an algorithm based on track segments, defined within the two innermost ITS layers, to detect multiple interaction vertices~\cite{ALICE:2014sbx}.~The influence of potentially remaining pile-up events is on the percent level and does not have an impact on the final results of the presented analysis.~After these selections, the data sample consisted of about 600 million events, corresponding to an integrated luminosity \mbox{$\Lint = 292 \pm 11~\upmu\text{b}^{-1}$}~\cite{ALICE:2014gvw}.~During the \pPb data taking period, the beam energies were $4$ TeV for protons and $1.58$ TeV per nucleon for lead nuclei. With this beam configuration, the nucleon--nucleon center-of-mass system moves in rapidity by $\Delta \ycms = 0.465$ in the direction of the proton beam. The charm-hadron analyses were performed in the laboratory-frame interval $|\ylab|<0.5$, leading to a shifted center-of-mass rapidity coverage of $-0.96 < \ycms < 0.04$.

\section{Analysis technique}
\label{analysisTech}

\subsection{Non-prompt \texorpdfstring{\Dzero}{D0}, \texorpdfstring{\Dplus}{D+}, and \texorpdfstring{\Lc}{Lc} raw yields}
\label{sec:analysisTech_sub1}

\begin{sloppypar}
The \Dzero, \Dplus, and \Lc charm hadrons, along with their charge conjugates, were reconstructed via the following hadronic decay channels: $\Dzero \to \kam \pi^+$  with branching ratio BR $= (3.95 \pm 0.03)\%$,
\mbox{$\Dplus \to  \pi^+\kam\pi^+$} with BR $= (9.38 \pm 0.16)\%$, $\Lc \to \proton \kam \pi^+$ with BR $= (6.28 \pm 0.32)\%$, and $\Lc \to \proton {\rm K^0_S} $ with ${\text{BR}=(1.59 \pm 0.08)\%}$, followed by $\mathrm{K^{0}_{S}} \to \pi^+\pi^-$ with BR $=(69.20 \pm 0.05)\%$)~\cite{ParticleDataGroup:2022pth}. The \Dzero-, \Dplus-, and \Lc-hadron candidates were defined by combining pairs or triplets of tracks reconstructed with the proper charge sign. While for the $\Lc \to \mathrm{pK_{s}^{0}}$ candidates, the V-shaped decay of the  $\mathrm{K_{\rm s}^{0}}$ meson into two pion-track candidates was combined with a proton-track candidate using a Kalman-Filter vertexing algorithm~\cite{kfparticle}, as described in~\cite{ALICE:2022exq}.~All daughter tracks were required to be reconstructed within $|\eta|<0.8$, with at least 70 associated space points in the \TPC, $\chi^2 / \rm{ndf} < 2$ of the fit quality of the \TPC tracks (where ndf is the number of degrees of freedom involved in the track fit procedure), and a minimum of 2 (out of 6) reconstructed clusters in the \ITS, with at least one in either of the two innermost layers.~These track-selection criteria reduce the D-meson and \Lc-baryon acceptance in rapidity, which drops steeply to zero for $|y_{\text{lab}}|>0.5$ at low \pt and for $|y_{\text{lab}}|>0.8$ at high \pt. Therefore, a \pt-dependent fiducial acceptance region $|y_{\text{lab}}| < y_{\text{fid}}(\pt)$ was applied to grant a uniform acceptance in the considered rapidity range.~The $y_{\text{fid}}(\pt)$ was defined as a second-order polynomial function, increasing from $0.5$ to $0.8$ in the transverse momentum range $0<\pt<5$ \GeVc, and as a constant term, $y_{\text{fid}} = 0.8$, for $\pt > 5 $ \GeVc.
\end{sloppypar}

The \Dzero, \Dplus, and \Lc decay weakly with a mean proper decay length ($c\tau$) of about $123$, $312$, and $60$~$\upmu$m, respectively~\cite{ParticleDataGroup:2022pth}. 
Charm hadrons coming from beauty-hadron decays are even more displaced from the primary vertex since their estimated $c\tau$ is about $500$~$\upmu$m, as for beauty hadrons. Therefore, these analyses were based on the reconstruction of decay-vertex topologies displaced from the primary vertex and, according to the selection applied, it is possible not only to separate candidates from the combinatorial background, but also the contributions of prompt and non-prompt charm hadrons.

To reduce the large combinatorial background and to separate the contributions of prompt and non-prompt charm hadrons, a machine-learning approach with multi-class classification, based on Boosted Decision Trees (BDT), implemented in the XGBoost library~\cite{Chen:2016btl,hipe4ml}, was adopted. For the BDT training, signal samples of prompt and non-prompt charm hadrons were obtained from simulations using the \textsc{PYTHIA} 8 event generator~\cite{Sjostrand:2014zea} (Monash-13 tune~\cite{Skands:2014pea}), embedded in an underlying \pPb collision generated with \textsc{HIJING} 1.36~\cite{wang1991hijing}, to describe better the charged-particle multiplicity and detector occupancy observed in the data.~Background samples were extracted from candidate invariant-mass distributions within the range of $5\sigma < |\Delta M| < 9\sigma$ in the data, where $\Delta M$ represents the difference between the candidate invariant mass and the nominal mass of the hadron candidate, and $\sigma$ represents the invariant-mass resolution. 

Before the training, loose selections were applied based on the decay kinematics and topologies along with the PID information of the decay-product tracks.~The PID selections were based on the difference between the measured and expected detector signals for a given particle species hypothesis, in units of the detector resolution ($n^\mathrm{det}_\sigma$). Protons, pions, and kaons were selected by requiring compatibility with the respective hypothesis within three standard deviations ($3\sigma$) for both the \TPC specific energy loss and the \TOF time-of-flight.~For tracks without a measured signal in the \TOF, the PID selections relied only on information from the \TPC.

Independent BDT models were trained for each \pt interval of the analysis of each D-meson species and the two \Lc decay channels using different variables related to the displaced decay-vertex topology and the PID information of the decay tracks.~The main variables used were (i) the distance of closest approach between the reconstructed tracks and the primary vertex, (ii) the distance between the charm-hadron decay vertex and the primary vertex, (iii) the charm-hadron impact parameter, (iv) the cosine of the pointing angle between the charm-hadron candidate line-of-flight and its reconstructed momentum. In the case of the $\Lc \to \proton {\rm K^0_S} $ decay, additional training variables related to the decay topology of the $\rm K^0_S$ and \Lc were used as in~\cite{ALICE:2022exq}.~The three BDT output scores are related to the candidate probability of being a prompt charm hadron, a non-prompt charm hadron, or combinatorial background. Selections on the non-prompt and combinatorial background BDT scores, corresponding to a requirement of a low probability for a candidate to be combinatorial background and a high probability to be non-prompt, were optimised to obtain a high non-prompt charm-hadron fraction in the inclusive signals while maintaining a reliable signal extraction, meaning a statistical significance larger than 3, as done in~\cite{ALICE:2021mgk, ALICE:2023brx}.

\begin{figure}[ht]
    \centering
    \includegraphics[keepaspectratio,width=0.45\linewidth]{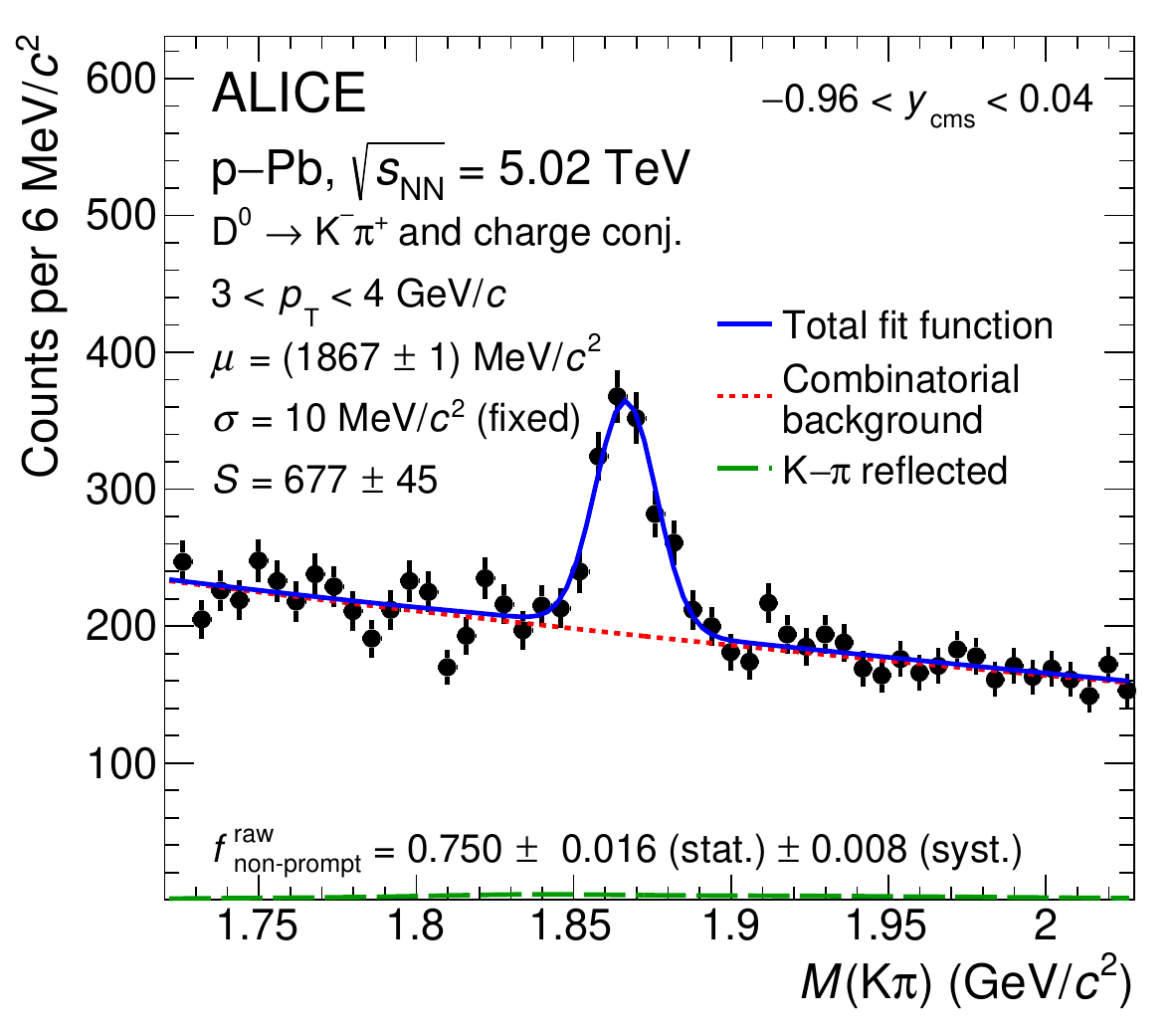}
    \includegraphics[keepaspectratio,width=0.45\linewidth]{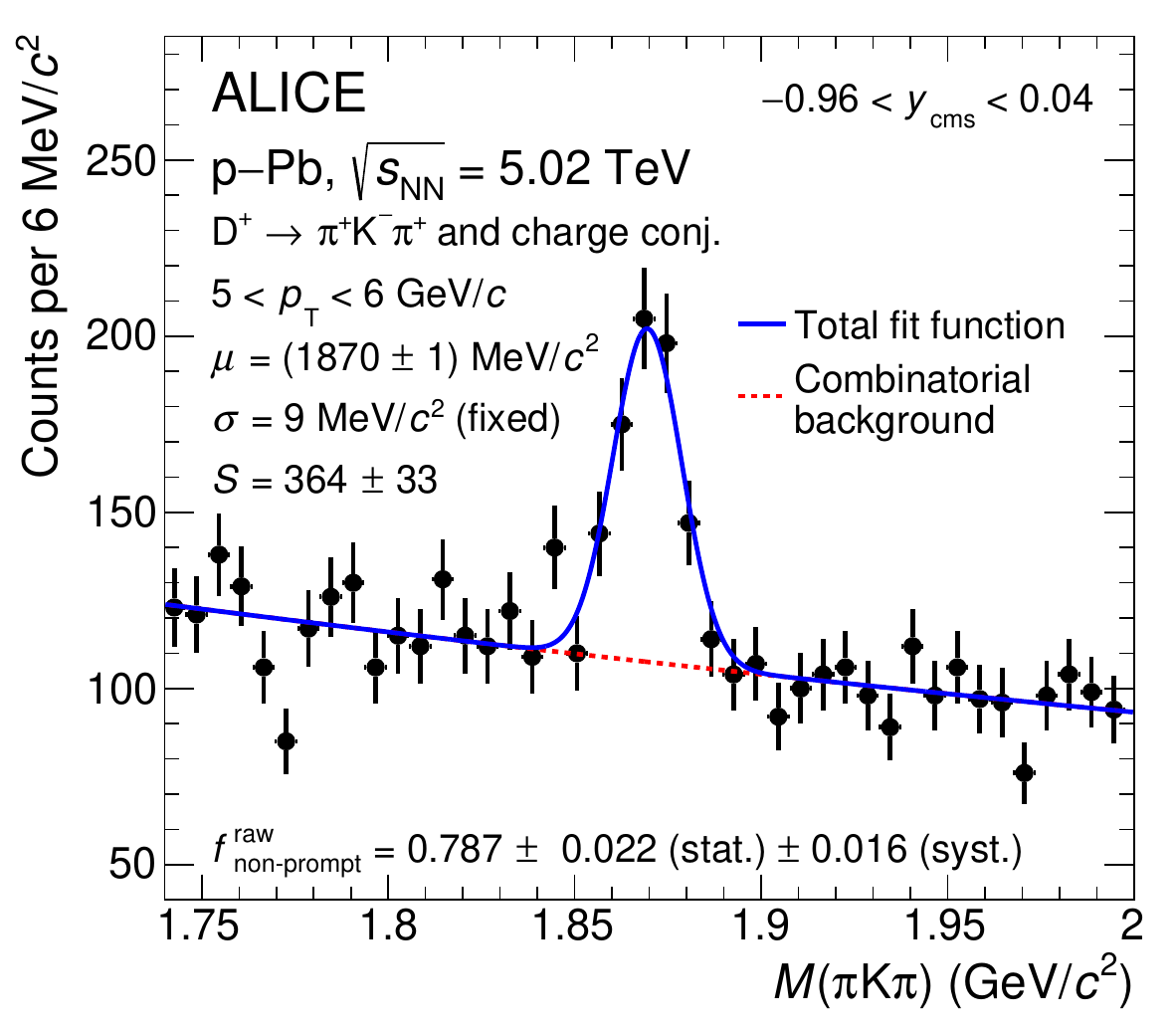}
    \includegraphics[keepaspectratio,width=0.45\linewidth]{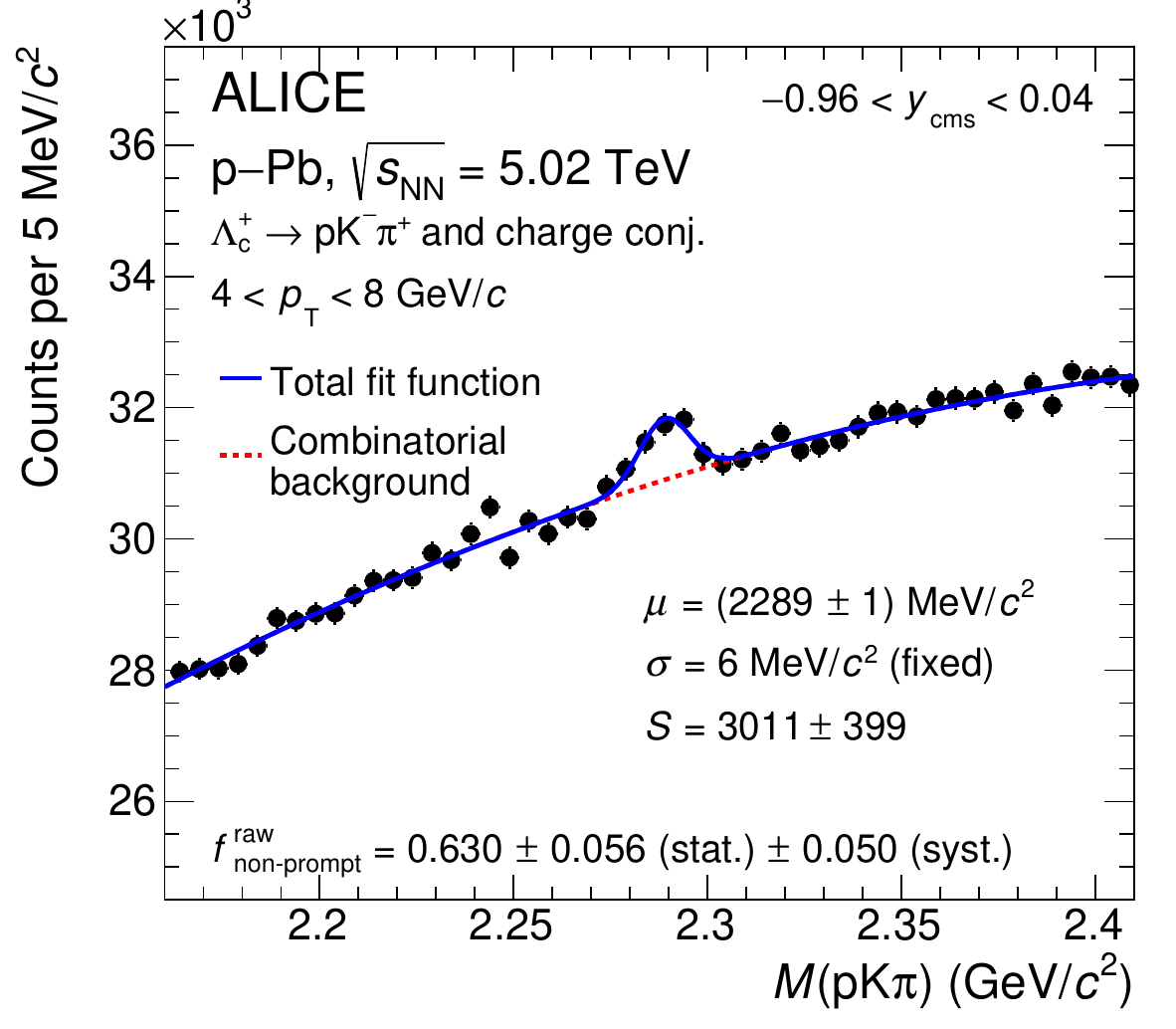}
    \includegraphics[keepaspectratio,width=0.45\linewidth]{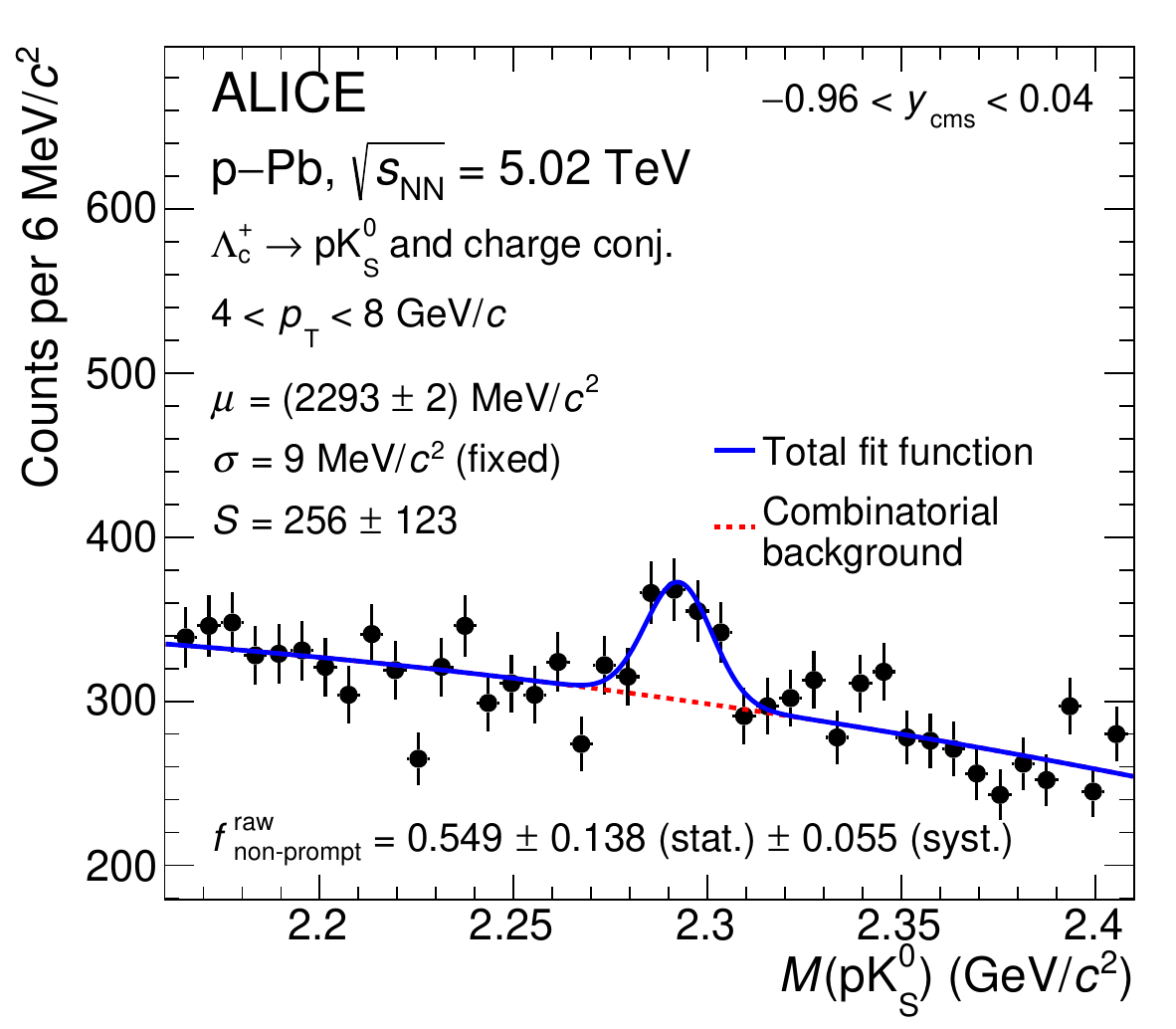}
    \caption{Invariant-mass distributions of \Dzero-, \Dplus-, and \Lc-hadron candidates, and their charge conjugates in selected \pt intervals. The blue solid lines show the total fit functions as described in the text and the red dashed lines show the fit function describing the combinatorial background. For \Dzero-meson candidates, the solid green line represents the contribution of the reflections. The mean ($\mu$) and fixed standard deviation ($\sigma$) of the signal fit function, along with the raw-yield ($S$) values, are reported together with their statistical uncertainties resulting from the fit. The fraction of non-prompt candidates in the measured raw yield ($f_{\text{non-prompt}}^{\mathrm{raw}}$) is reported with its statistical and systematic uncertainties.} 
    \label{fig:invariant_mass}
\end{figure}

The raw yields of \Dzero, \Dplus, and \Lc hadrons, including particles and antiparticles, were extracted via binned maximum-likelihood fits to the invariant-mass ($M$) distributions of the selected charm-hadron candidates. The raw yields were extracted in transverse-momentum intervals in the range $1 < \pt < 24$~\GeVc for \Dzero mesons, and $2 < \pt < 24$~\GeVc for \Dplus mesons and \Lc baryons. The fitting function was composed of an exponential or polynomial term for describing the background and a Gaussian term for the signal. To improve the stability of the fits, the widths of the charm-hadron signal peaks were fixed to the values extracted from data samples enhanced with prompt candidates, given the naturally larger abundance of prompt compared to non-prompt charm hadrons. As part of the systematic uncertainty analysis, the width parameter was varied to determine its impact on the systematic uncertainty associated with the raw-yield extraction. For the \Dzero mesons, the contribution of signal candidates to the invariant-mass distribution with the wrong mass assigned to the \Dzero-decay tracks, referred to as reflections, was included in the fit, and estimated as explained in~\cite{ALICE:2023brx}. The contribution of reflections to the raw yield is about 1--2\%, depending on \pt. Examples of invariant-mass distributions together with the result of the fits and the estimated non-prompt fractions ($f_{\text{non-prompt}}^{\mathrm{raw}}$) are reported in Fig.~\ref{fig:invariant_mass}, for the $3 < \pt < 4$~\GeVc, $5 < \pt < 6$~\GeVc, and $4 < \pt < 8$~\GeVc intervals of the \Dzero, \Dplus, and \Lc hadrons, respectively.

\subsection{Yield corrections and non-prompt fraction estimations}
\label{sec:non_prompt_fraction}

The \pt-differential production cross sections of non-prompt \Dzero, \Dplus, and \Lc hadrons at midrapidity were computed as:
\begin{equation}
    \label{eq:cross_section}
    \left. \cfrac{\mathrm{d}^2 \sigma^{\mathrm{H}_{\mathrm{c}}}}{ \mathrm{d}  \pt \mathrm{d}y} \right|_{|\ylab| < 0.5} = 
    \cfrac{1}{2} \times  
    \cfrac{ f_{\text{non-prompt}}^{\text{raw}} (\pt) \times   \left. N^{\mathrm{H}_{\mathrm{c}} + \overline{\mathrm{H}}_{\mathrm{c}}, \mathrm{raw}} (\pt) \right|_{|y_{\mathrm{lab}}|<y_{\text{fid}} (\pt)} }{ \Delta \pt \times  c_{\Delta y}(\pt)  \times (\mathrm{Acc} \times \varepsilon)^{\text{non-prompt}} (\pt)} \cfrac{1}{\mathrm{BR} \times \mathcal{L}_{\mathrm{int}}} ~,
\end{equation}
where $N^{\mathrm{H}_{\mathrm{c}} + \overline{\mathrm{H}}_{\mathrm{c}}, \mathrm{raw}}$ (sum of particles and antiparticles) represents the raw yields extracted in each \pt interval, and the factor 1/2 is included to account that the raw yields contain both particles and antiparticles, while the production cross sections are given as an average of particles and antiparticles. The $f_{\text{non-prompt}}^{\text{raw}}$ factor represents the raw non-prompt fraction needed to account for the residual contribution of prompt charm hadrons in the extracted non-prompt raw yields. In addition, the yields were further divided by the width of the \pt interval ($\Delta \pt$), the correction factor for the rapidity coverage $c_{\Delta y}$, computed as the ratio between the generated hadron yield in $ \Delta y = 2 \yfid$ and that in $|\ylab | < 0.5$, as explained in~\cite{ALICE:2019fhe}, as well as the acceptance times efficiency of non-prompt charm hadrons $(\mathrm{Acc} \times \varepsilon)^{\text{non-prompt}}$, the BR of the decay channel, and the integrated luminosity $\Lint$.

\begin{figure}[ht]
    \centering
    \includegraphics[keepaspectratio,width=0.45\linewidth]{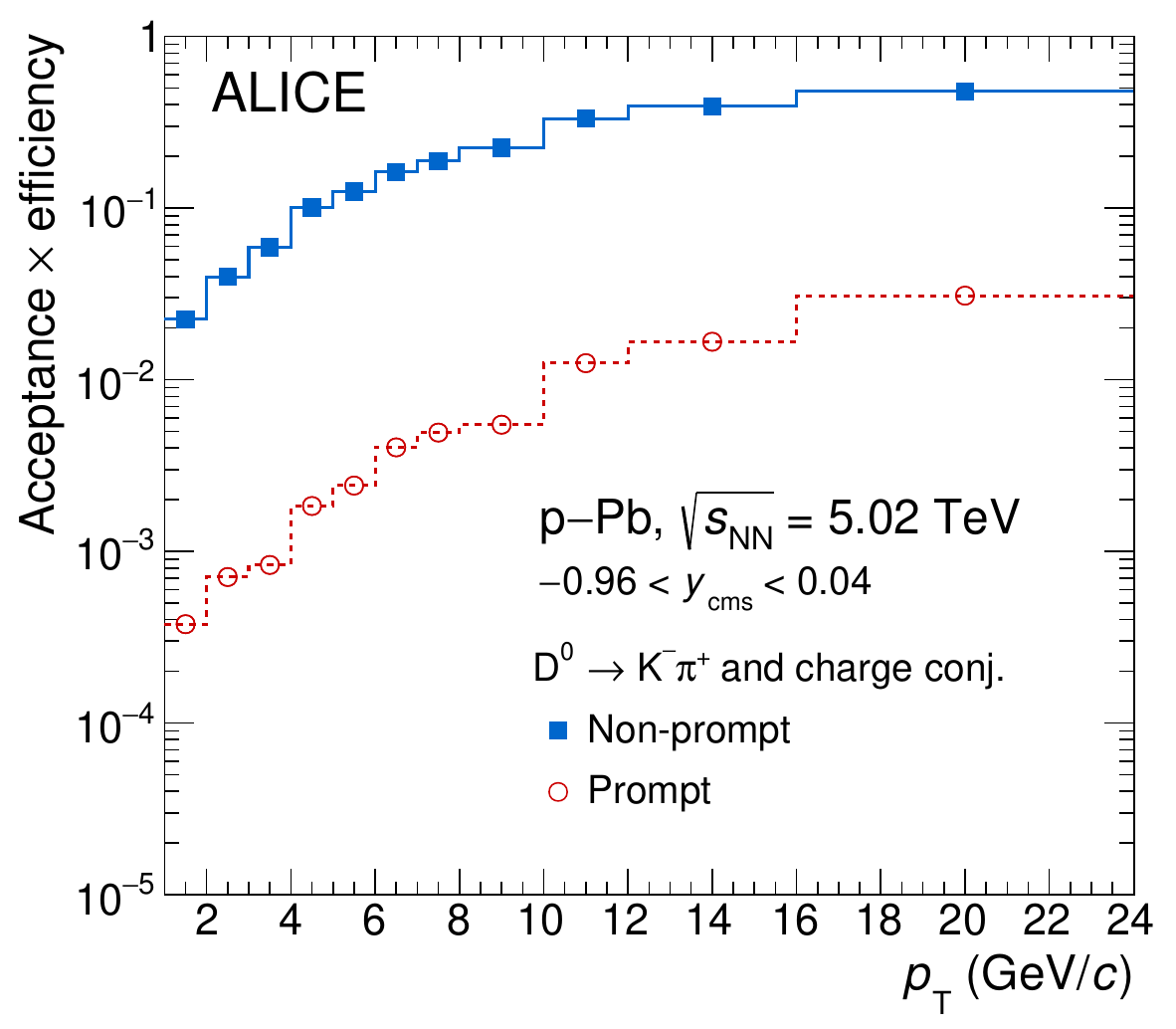}
    \includegraphics[keepaspectratio,width=0.45\linewidth]{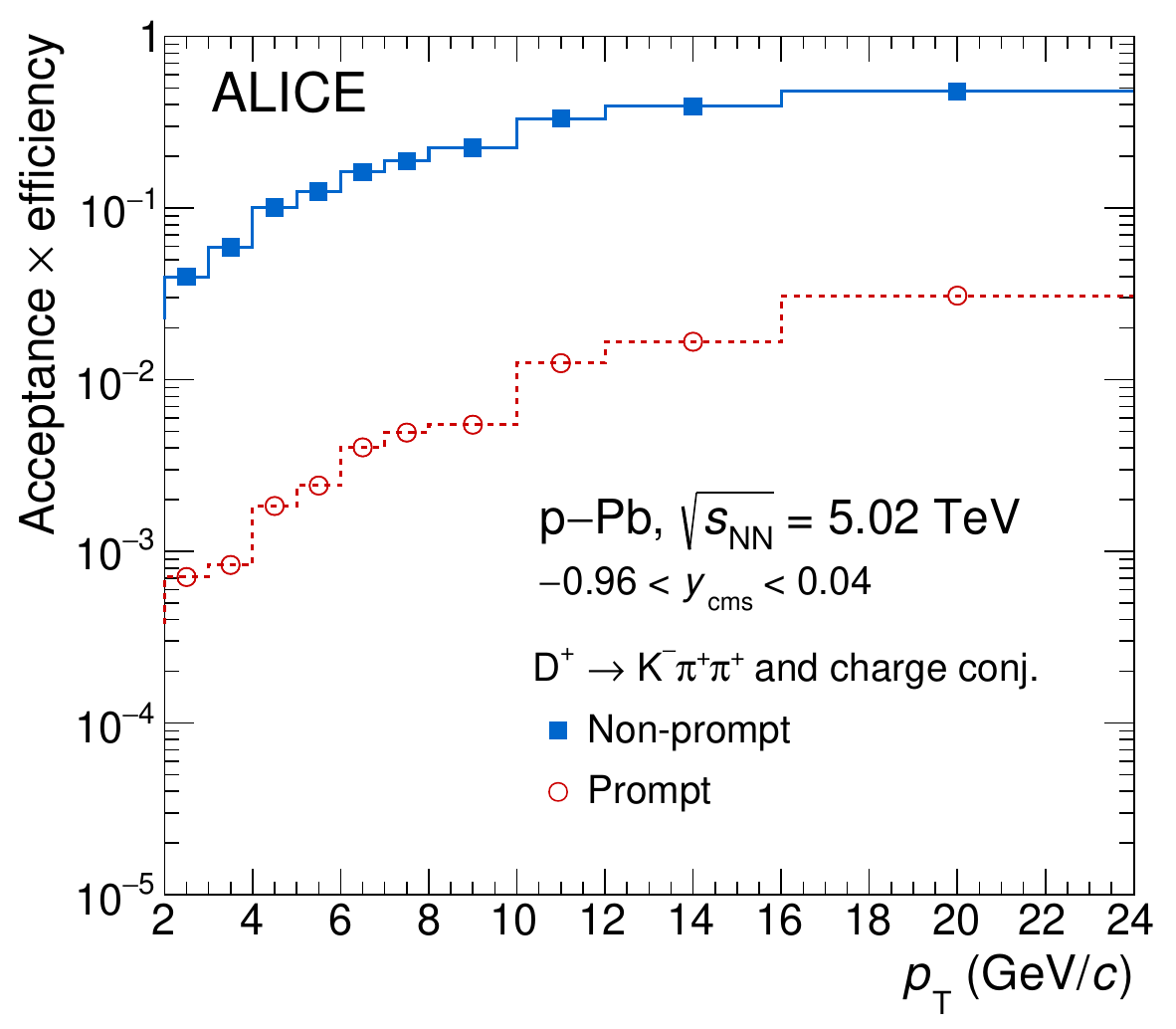}
    \includegraphics[keepaspectratio,width=0.45\linewidth]{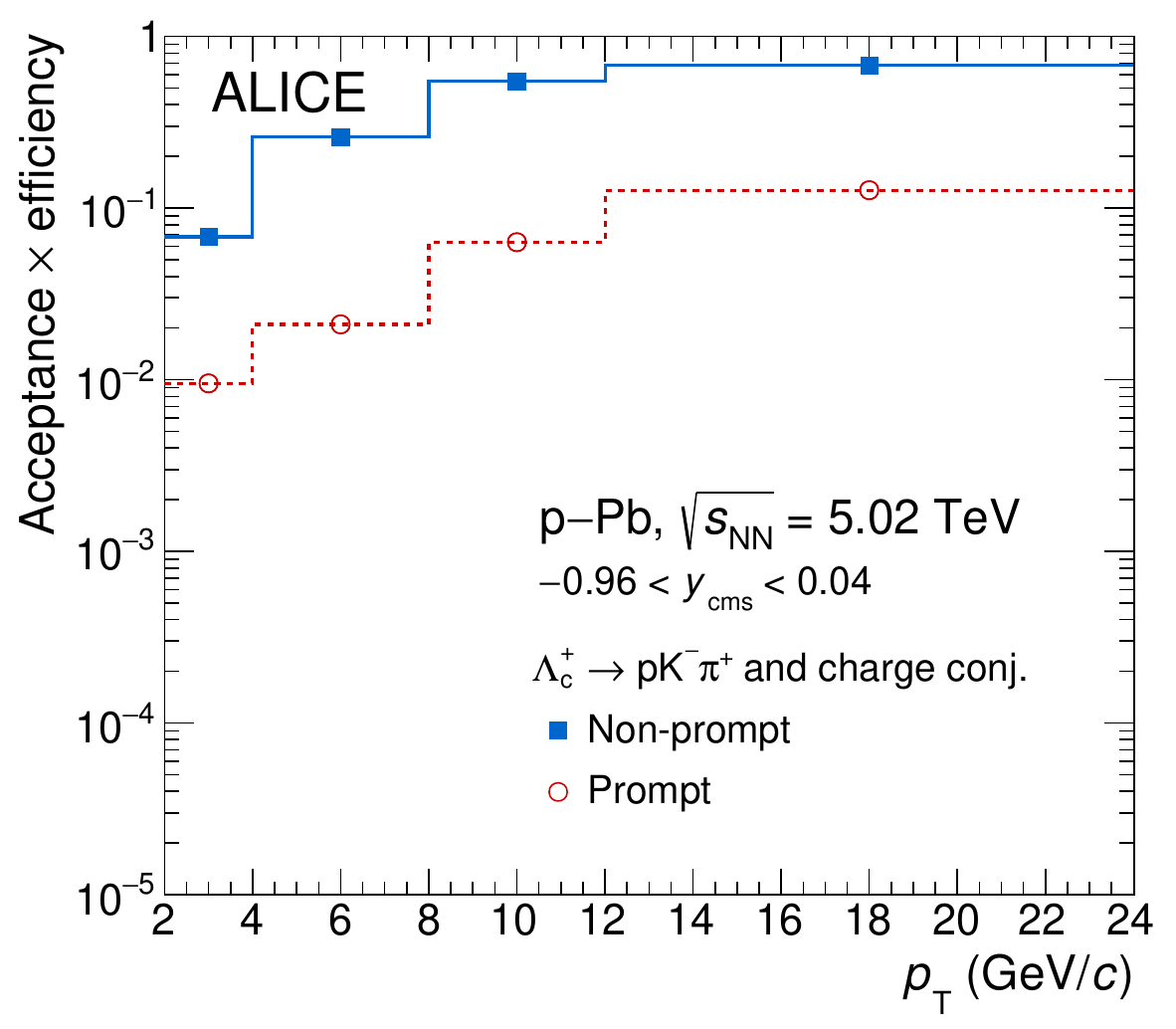}
    \includegraphics[keepaspectratio,width=0.45\linewidth]{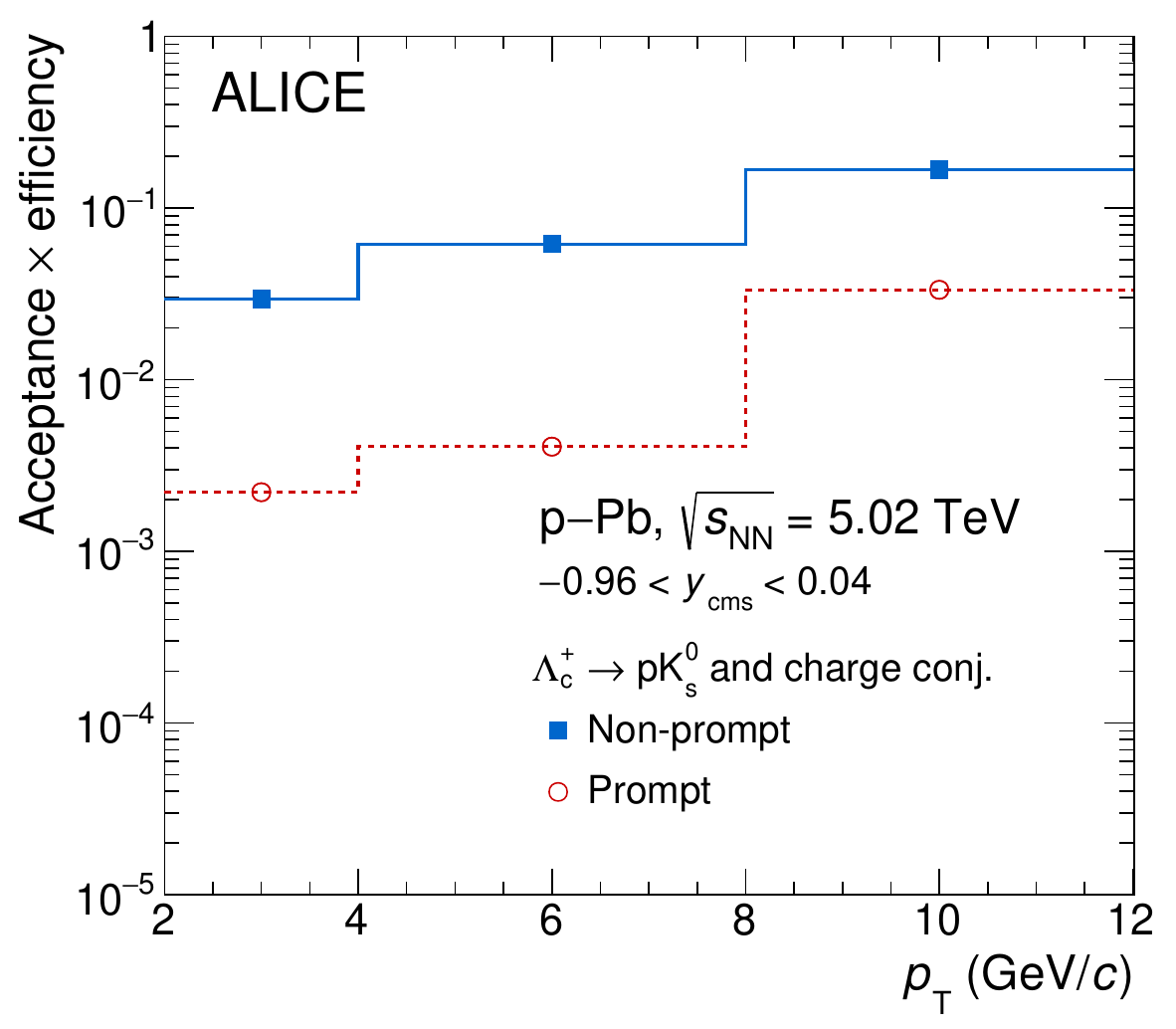}
    \caption{Acceptance-times-efficiency factors for \Dzero, \Dplus, and \Lc hadrons as a function of \pt. }
    \label{fig:acceptance_times_efficiency}
\end{figure}

Possible differences in the \pt shape of prompt and non-prompt charm hadrons between data and Monte Carlo (MC) simulations were corrected by weighting the simulated \pt distribution of prompt charm hadrons and of the beauty-hadron parent, respectively. For D mesons, the weights were computed by dividing the \pt spectrum predicted by FONLL calculations and the one obtained from \textsc{PYTHIA}~8 simulations. The FONLL \pt spectra of prompt and non-prompt D mesons in p--Pb collisions were computed using the predictions in \pp collisions at $\s = 5.02$ TeV~\cite{Cacciari:1998it, Cacciari:2001td}, assuming that the \RpPb of D and B mesons are compatible with unity, in the rapidity range of this study.~This assumption is based on the D-meson \RpPb measurements at \snn = 5.02 TeV at midrapidity by the ALICE Collaboration~\cite{ALICE:2019fhe} and B-meson \RpPb measurements at \snn = 8.16 TeV at forward/backward rapidity by the LHCb Collaboration~\cite{LHCb:2019avm}, that are in agreement with models that predict B-meson \RpPb values compatible with unity at midrapidity, within the theoretical uncertainties. The energy dependence of the \RpPb measurement is neglected. The procedure to compute the \pt spectrum based on FONLL calculations for the prompt \Lc (\Lb) hadrons takes into account three essential components: the FONLL predicted \pt distribution for prompt \Dzero (B) mesons in pp collisions at $\s = 5.02$ TeV, the prompt \Lc/\Dzero  ($\Lb/\mathrm{B}^{0}$) ratio~\cite{ALICE:2022exq, LHCb:2019fns} in the same collision system and energy, and the prompt \Lc \RpPb at $\snn = 5.02$ TeV measured by the ALICE Collaboration~\cite{ALICE:2022exq}. The weights on prompt \Lc (\Lb) \pt shape were derived as the product of these three components divided by the prompt \Lc (\Lb) MC \pt distribution from \textsc{PYTHIA} 8.~In addition, the weights on non-prompt \Lc were derived from \Lb based on the \pt correlation between \Lb and non-prompt \Lc from \Lb decays simulated by \textsc{PYTHIA} 8.

The $(\mathrm{Acc} \times \varepsilon)$ correction was obtained from simulations, as described in Section~\ref{sec:analysisTech_sub1}, using samples not employed for the BDT training.~The $(\mathrm{Acc} \times \varepsilon)$ factors, computed for non-prompt \Dzero, \Dplus, and \Lc hadrons as a function of \pt, after applying all the selections, are shown in Fig.~\ref{fig:acceptance_times_efficiency}. The selection applied to obtain the non-prompt enhanced samples strongly suppresses the prompt charm-hadron efficiency. The prompt charm-hadron acceptance-times-efficiency $(\mathrm{Acc} \times \varepsilon)$ is smaller than the non-prompt one by factors varying from 20 to 60 for D mesons and 5 to 13 for \Lc baryons, depending on the \pt interval.

A data-driven procedure, based on the construction of data samples with different abundances of prompt and non-prompt candidates, was used to estimate the fraction $f_{\text{non-prompt}}^{\mathrm{raw}}$ of non-prompt \Dzero, \Dplus, and \Lc hadrons in the extracted yields.~Let $i \in \{ 1; n\}$ designate a set among $n \in \mathbb{N}$ selection sets.~Each set of BDT selection $i$ is associated with an extracted raw-yields value ($Y_i$), which relates to the corrected yield of prompt ($N_{\mathrm{prompt}}$) and non-prompt ($N_{\mathrm{non-prompt}}$) charm hadrons via the corresponding prompt $(\mathrm{Acc} \times \varepsilon)_{i}^{\mathrm{prompt}}$ and non-prompt $(\mathrm{Acc} \times \varepsilon)_{i}^{\mathrm{non-prompt}}$ efficiency as follows:

\begin{equation}
\label{eq:2}
    (\mathrm{Acc} \times \varepsilon)^{\rm prompt}_i \times N_{\rm prompt} + (\mathrm{Acc} \times \varepsilon)^{\text{non-prompt}}_i \times N_{\text{non-prompt}} - Y_i = \delta_i ~,
\end{equation}

where $\delta_i$ represents a residual that accounts for the equation not holding precisely due to the uncertainties of $Y_i$, $(\mathrm{Acc} \times \varepsilon)^{\text{non-prompt}}_i$, and $(\mathrm{Acc} \times \varepsilon)^{\rm prompt}_i$. In the case of $n \geqslant$ 2 sets, a $\chi^2$ function can be defined based on Eq.~\ref{eq:2}, which can be minimised to obtain $N_{\rm prompt}$ and $N_{\text{non-prompt}}$ as explained in ~\cite{ALICE:2023wbx,ALICE:2021mgk}. 

\begin{figure}[ht]
    \centering
    \includegraphics[keepaspectratio,width=0.48\linewidth]{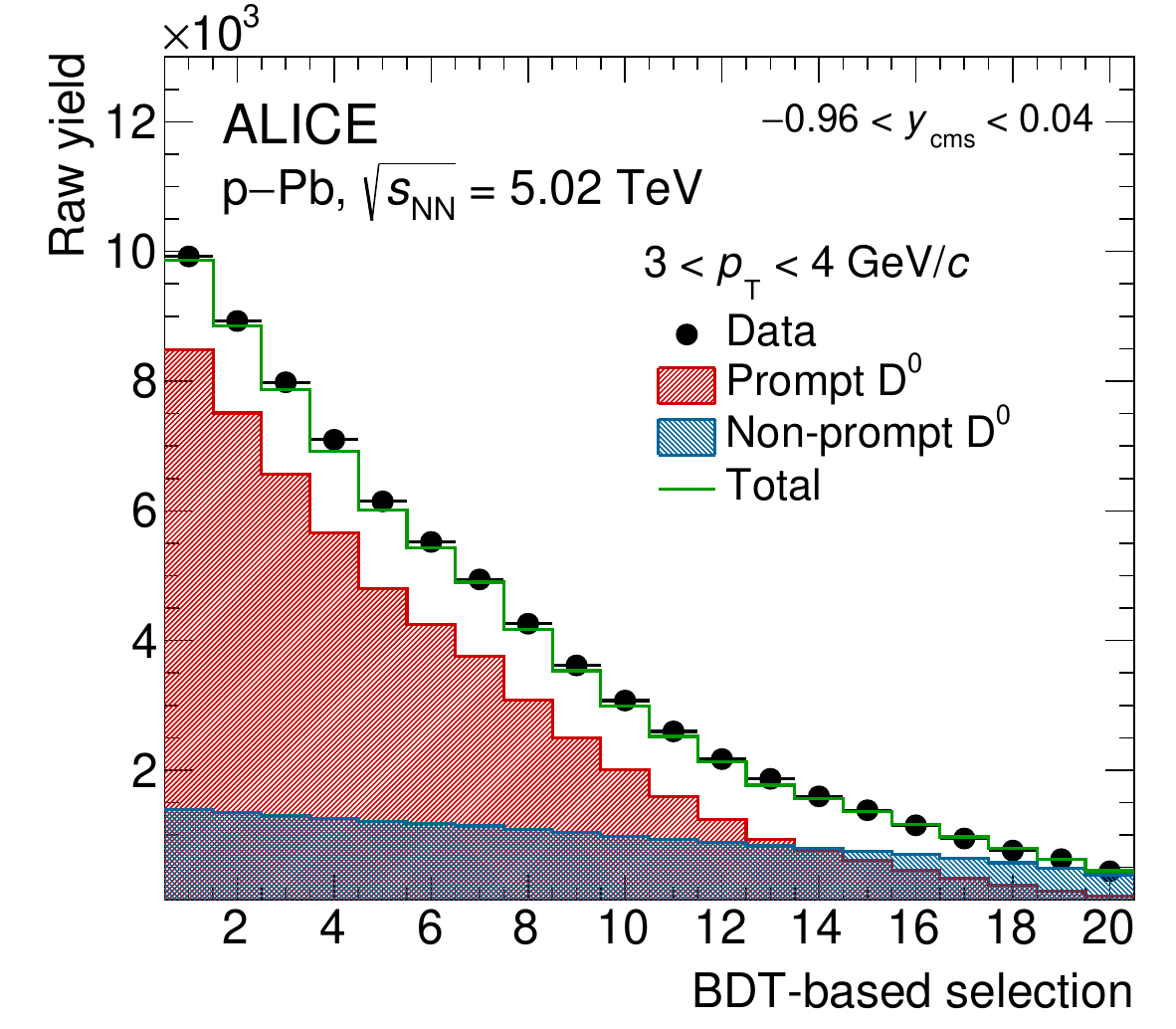}
    \includegraphics[keepaspectratio,width=0.48\linewidth]{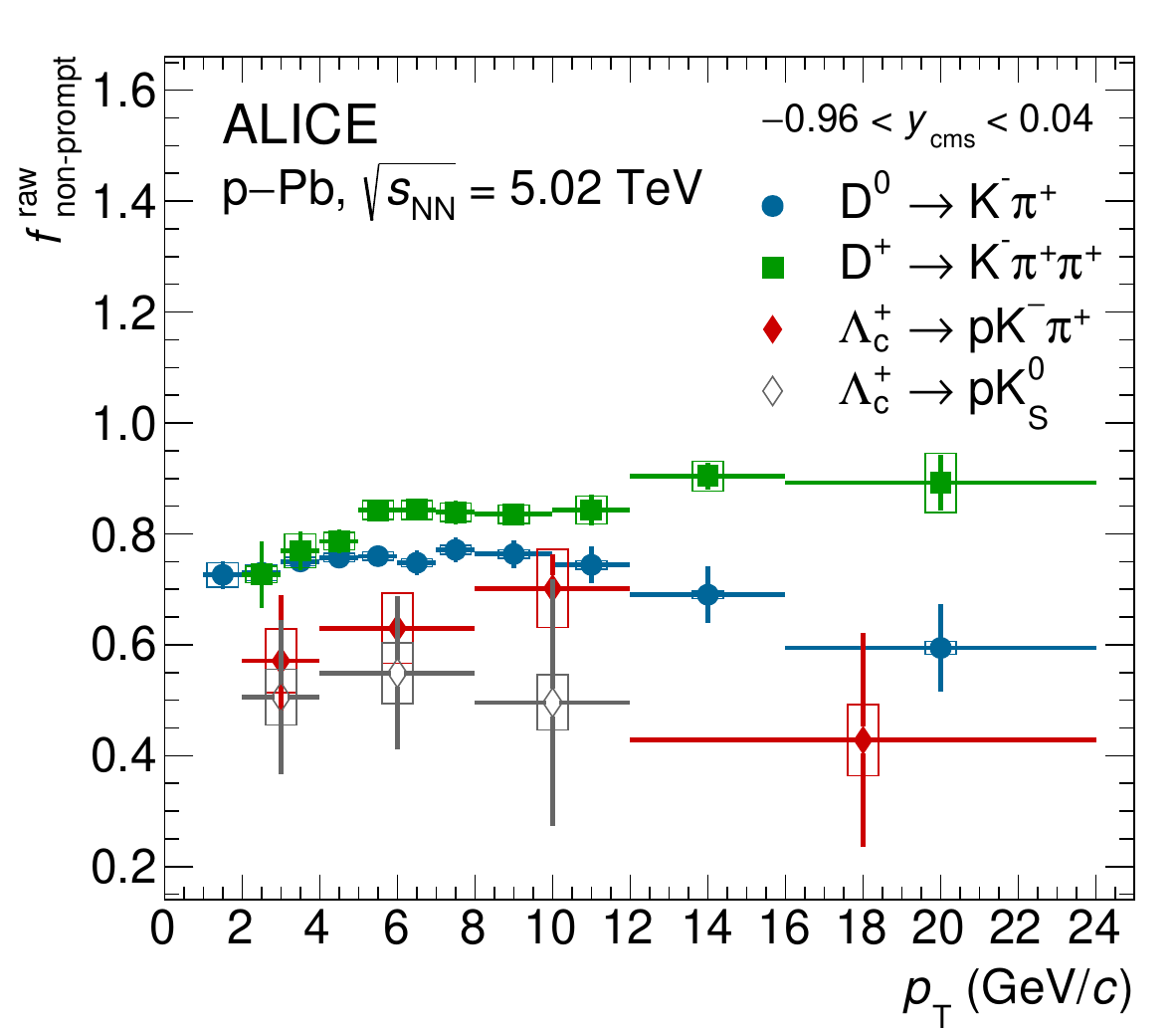}
    \caption{Left panel: example of raw-yield distribution as a function of the BDT-based selection employed in the data-driven procedure adopted to determine $f_{\text{non-prompt}}^{\mathrm{raw}}$ of \Dzero mesons. Right panel: $f_{\text{non-prompt}}^{\mathrm{raw}}$ fractions as a function of \pt obtained for the set of selection criteria adopted in the analysis for non-prompt \Dzero, \Dplus, $\Lc \to \mathrm{pK^{-}\pi^{+}}$, and $\Lc \to \mathrm{pK^{0}_{S}}$. The vertical bars and empty boxes represent the statistical and systematic uncertainties, respectively.}
    \label{fig:cut_variation_method}
\end{figure}

Figure~\ref{fig:cut_variation_method} shows an example of the raw-yield distribution as a function of the BDT-based selection employed in the minimisation procedure for \Dzero mesons in $3 < \pt < 4$ \GeVc (left panel).~The leftmost data point of the distribution is the raw yield corresponding to the looser selections on the BDT outputs related to the candidate probability of being a non-prompt charm hadron. In contrast, the rightmost one corresponds to the tightest selections. The right panel shows the \pt distributions of the raw non-prompt fraction $f_{\text{non-prompt}}^{\mathrm{raw}}$ obtained for the set of selection criteria adopted in the analysis for non-prompt \Dzero, \Dplus, $\Lc \to \mathrm{pK^{-}\pi^{+}}$, and $\Lc \to \mathrm{pK^{0}_{S}}$. The fraction $f_{\text{non-prompt}}^{\mathrm{raw}}$ of \Dzero, \Dplus, and $\Lc$ hadrons ranges from 42 to 90\% depending on the decay channel and the \pt interval.

\section{Systematic uncertainties}
\label{systUnc}

The measurement of the \pt-differential production cross section of non-prompt charm hadrons was affected by the following sources of systematic uncertainties: (i) extraction of the raw yield from the invariant-mass distribution, (ii) non-prompt fraction estimation, (iii) corrections to the generated \pt shape in simulations, (iv) charm-hadron selection efficiency, and (v) track-reconstruction efficiency.~The systematic uncertainties of the PID selection efficiency were found to be negligible, as observed in prompt charm hadron measurements~\cite{ALICE:2019fhe,ALICE:2022exq}.~In addition, the \pt-differential production cross section was affected by the uncertainties on the branching ratios of the considered charm-hadron decays~\cite{ParticleDataGroup:2022pth} and a systematic uncertainty on the overall normalisation induced by the uncertainties on the integrated luminosity of $3.7\%$~\cite{ALICE:2014gvw}. The values of the systematic uncertainties for some representative \pt intervals were summarised in Table~\ref{tab:systematics}.~The contributions of the different sources were considered to be uncorrelated and were summed in quadrature to obtain the total systematic uncertainty.

\begin{table}[ht]
    \centering
    \caption{Summary of the relative systematic uncertainties on the measurement of non-prompt \Dzero, \Dplus, and \Lc production cross sections in different \pt intervals.}
    \begin{tabular}{|l|cc|cc|cc|cc|}
    \hline
Hadron & \multicolumn{2}{c|}{$\Dzero \left( \to \kam \pip \right)$} & \multicolumn{2}{c|}{$\Dplus \left(\to \pip \kam \pip \right)$} & \multicolumn{2}{c|}{$\Lc \left(\to \proton \kam \pip \right)$} &  \multicolumn{2}{c|}{$\Lc \left(\to \proton \Kzs \right)$} \\
\pt (\GeVc)          & 1--2    &  10--12   &  2--3    &  10--12   &  2--4    &  12--24   &  2--4    &   8--12   \\
\hline
Signal yield         &  3\%    &    2\%    &   6\%    &    5\%    &   7\%    &    15\%   &   10\%   &    9\%    \\
Fraction estimation  &  3\%    &    1\%    &   2\%    &    3\%    &   10\%   &    15\%   &   10\%   &    10\%   \\
\pt shape in MC      &  7\%    &    0\%    &   1\%    &    0\%    &   5\%    &    0\%    &   5\%    &    0\%    \\
Selection efficiency &  5\%    &    4\%    &   6\%    &    3\%    &   8\%    &    8\%    &   7\%    &    7\%    \\
Tracking efficiency  &  2.0\%  &    2.5\%  &   3.7\%  &    4.0\%  &   6.0\%  &    6.0\%  &   5.0\%  &    5.0\%  \\
Branching ratio~\cite{ParticleDataGroup:2022pth} & \multicolumn{2}{c|}{0.8\%} & \multicolumn{2}{c|}{1.7\%} & \multicolumn{2}{c|}{5.1\%} & \multicolumn{2}{c|}{5.0\%} \\
\hline
Luminosity~\cite{ALICE:2014gvw} & \multicolumn{8}{c|}{3.7\%} \\
\hline
    \end{tabular}
    \label{tab:systematics}
\end{table}

The systematic uncertainty on the raw yield extraction was evaluated for each charm-hadron species by repeating the fits to the invariant-mass distribution for each \pt interval of the analyses, varying the fit range, the functional form of the background fit function, the bin size of the invariant mass spectrum, and the width of the Gaussian function used to model the signal peaks. The latter was varied within the uncertainty of the value obtained from the fits to the prompt candidate enhanced data sample. The systematic uncertainty was defined as the root mean square of the distribution of the signal yields obtained from the described variations and ranged from 2 to $15\%$ depending on the hadron species and the \pt interval.

The systematic uncertainty on the value of $f_{\text{non-prompt}}^{\text{raw}}$ obtained with the data-driven approach was estimated by varying the number of BDT selections employed in the data-driven method as described in Section~\ref{sec:non_prompt_fraction}. A systematic uncertainty ranging from 1 to 15\% was assigned.

The systematic effect due to the dependence of the efficiencies on the generated \pt distribution of heavy-flavour hadrons was estimated by evaluating the production cross section after weighting the \pt shape of the PYTHIA 8 generator to match the central one predicted by FONLL calculations, as well as the upper- and lower-edge of the predictions which account for the uncertainties due to the choice of the heavy-quark masses, factorisation and renormalisation scales, and the uncertainties on the CTEQ6.6 PDFs~\cite{Pumplin:2002vw}. The weights were applied to the \pt distributions of prompt charm hadrons and of the beauty hadron parent in the case of non-prompt charm hadrons.~The assigned systematic uncertainty, considering the root mean square of the production cross section distributions obtained for minimal and maximal FONLL predictions with respect to the central (default) ones, reached up to $7\%$.

The systematic uncertainty on the selection efficiency originates from imperfections in the description of the kinematic and topological variables of the candidates and of the detector resolutions and alignments in the simulation. It was estimated by comparing the production cross sections obtained by repeating the analysis with different selections on the BDT outputs, resulting in a significant modification of the efficiency values. The assigned systematic uncertainty ranged from 3 to 8\%.

The systematic uncertainties on the track reconstruction efficiency were estimated by considering the uncertainty due to track quality selections and the uncertainty due to the TPC–ITS track matching efficiency as discussed in~\cite{ALICE:2019fhe,ALICE:2022exq}. It ranged from 2 to 6\%, depending on the candidate species and \pt interval.

\section{Results}
\label{result}

\subsection{Production cross sections}

\begin{figure}[ht]
    \centering
    \includegraphics[keepaspectratio,width=0.45\linewidth]{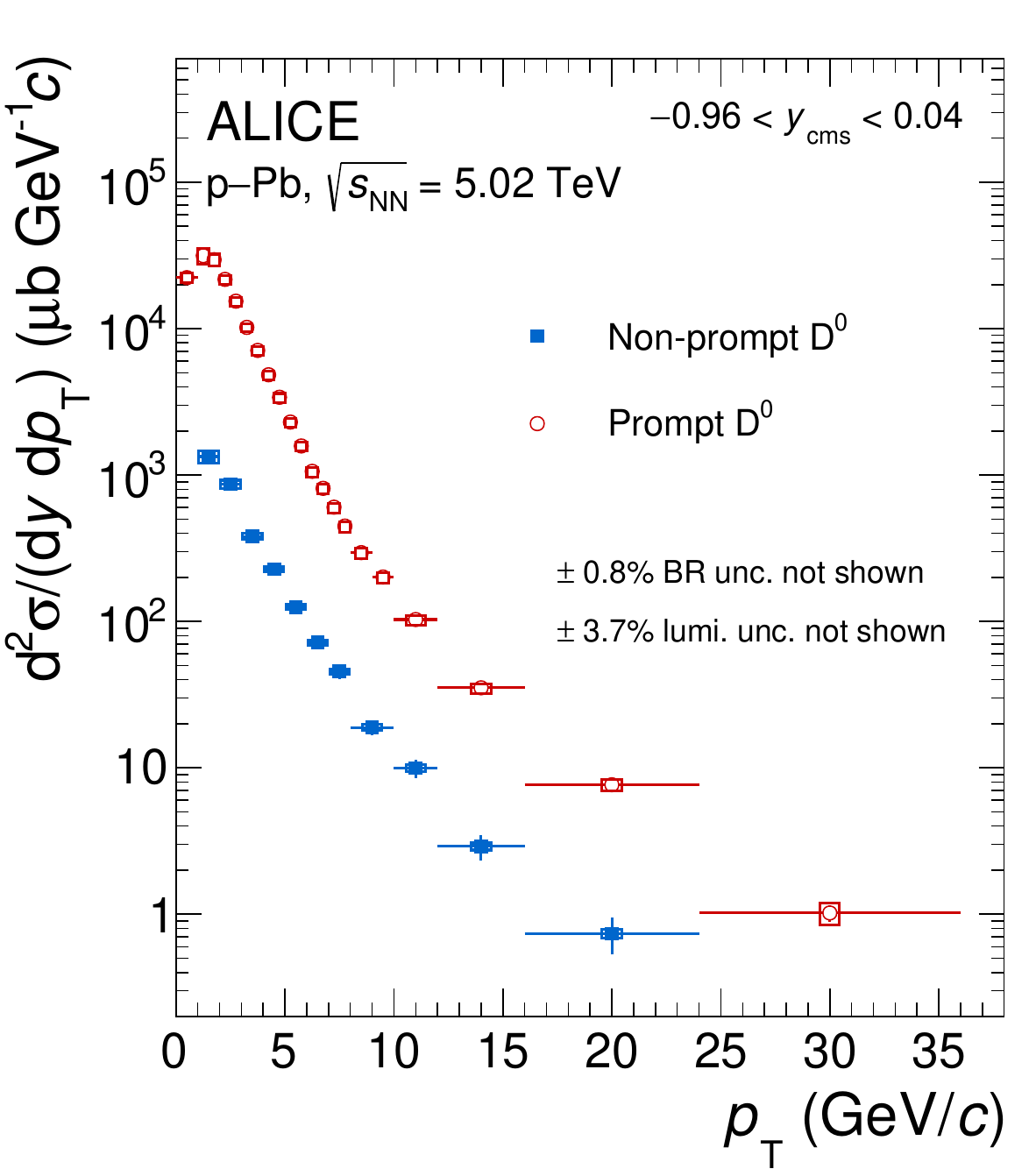}
    \includegraphics[keepaspectratio,width=0.45\linewidth]{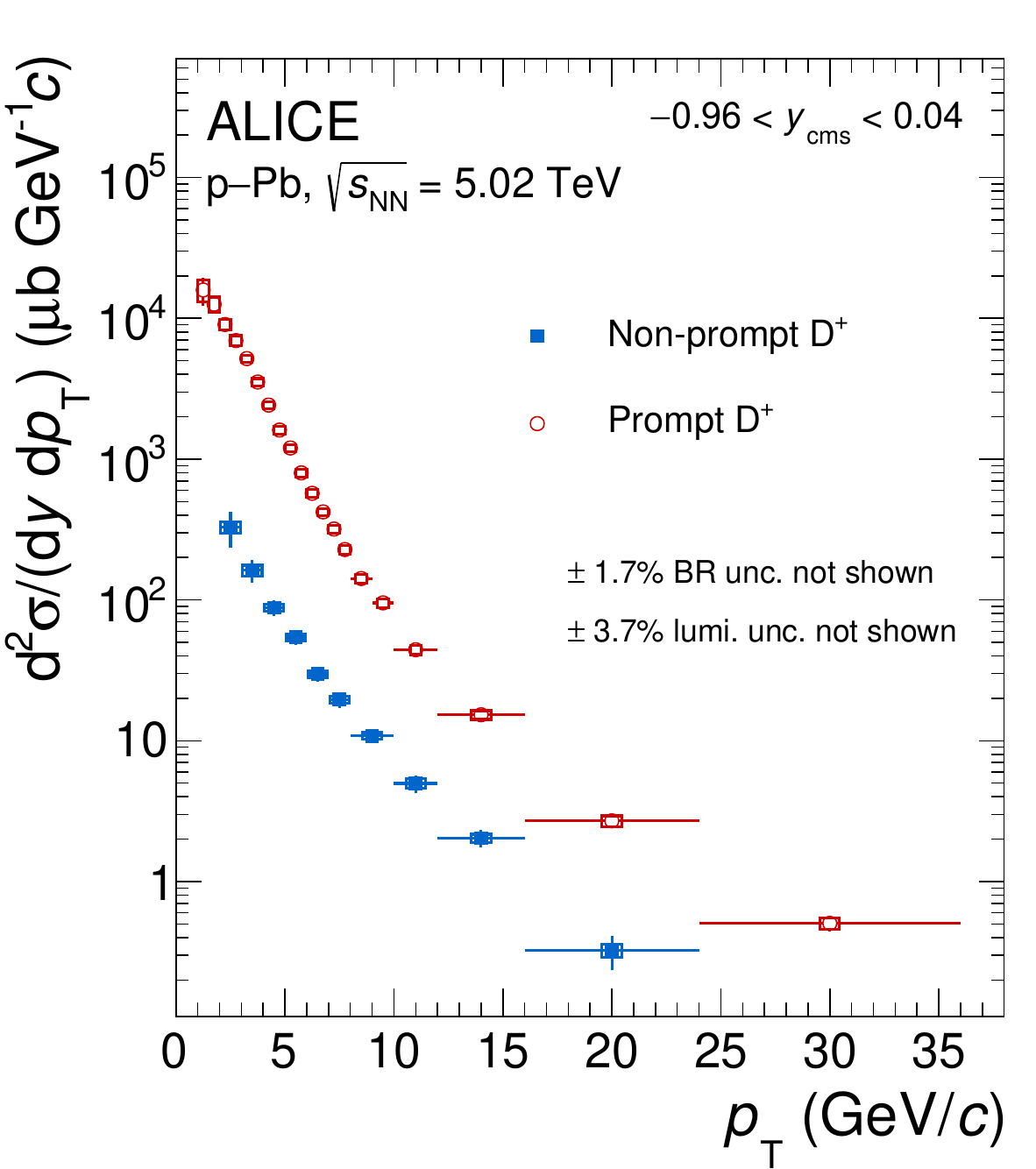}
    \includegraphics[keepaspectratio,width=0.45\linewidth]{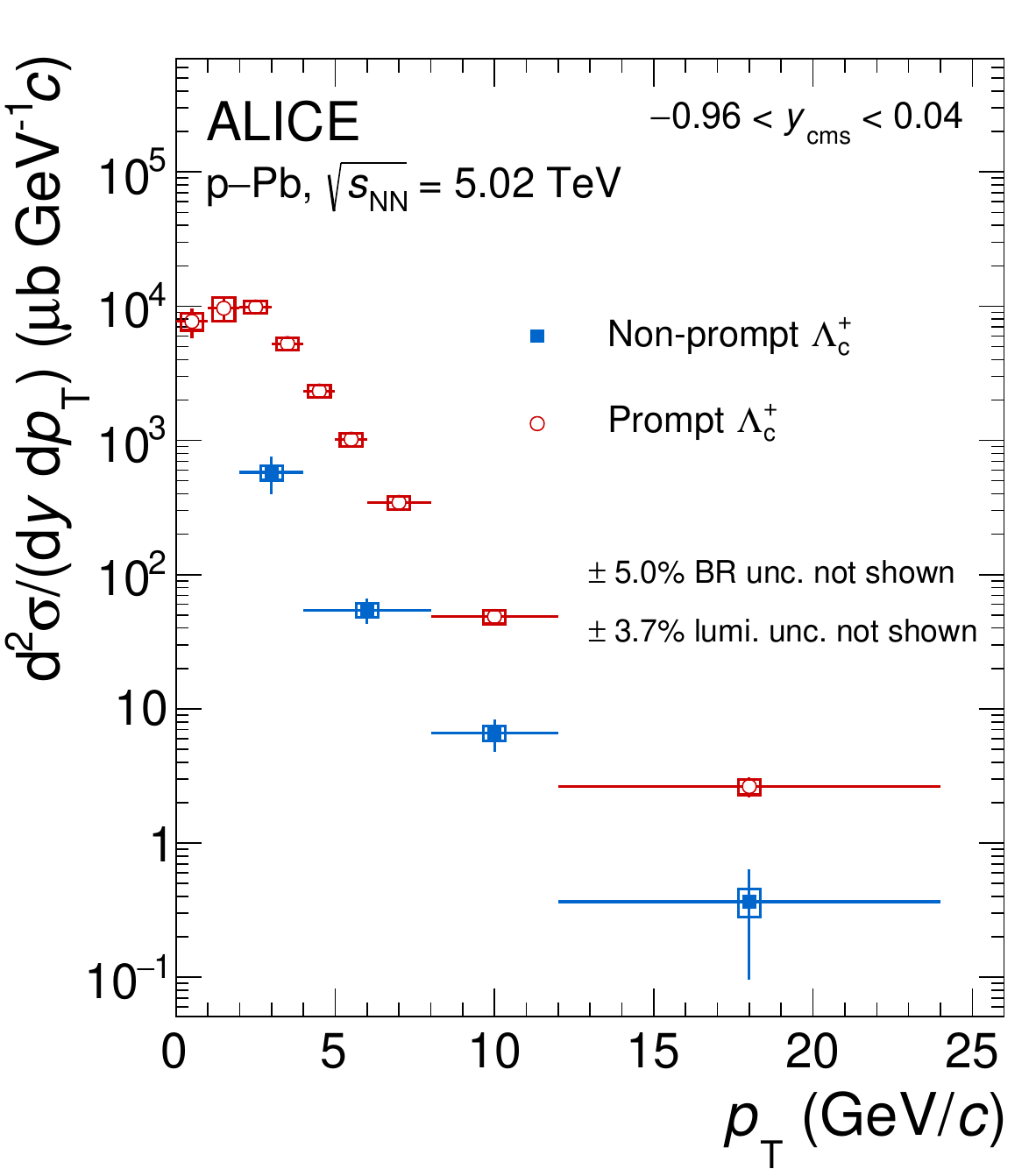}
    \caption{\pt-differential production cross sections of non-prompt \Dzero, \Dplus, and \Lc in \pPb collisions at $\snn = 5.02$ \TeV, in comparison with the corresponding production cross section of prompt hadrons from~\cite{ALICE:2019fhe, ALICE:2022exq}. The measurement of prompt \Dplus mesons is the one reported in~\cite{ALICE:2019fhe}, with decay BR discussed in the text.~The vertical bars and empty boxes represent the statistical and systematic uncertainties (without branching ratio and luminosity contributions), respectively.}
    \label{fig:CrossSectionD0DplusLambdac}
\end{figure}

The \pt-differential production cross sections of non-prompt \Dzero mesons, \Dplus mesons, and \Lc baryons in \pPb collisions at \snn = 5.02 TeV, measured in the rapidity interval $-0.96 < \ycms < 0.04 $, are shown in Fig.~\ref{fig:CrossSectionD0DplusLambdac} in comparison to those measured for prompt hadrons at the same center-of-mass energy~\cite{ALICE:2019fhe, ALICE:2022exq}. The measurement of prompt \Dplus is the one reported in~\cite{ALICE:2019fhe}, scaled for the BR $= (8.98 \pm 0.28)\%$ of the \mbox{$\Dplus \to  \pi^+\kam\pi^+$} decay reported in~\cite{ParticleDataGroup:2018ovx}. The non-prompt \Lc-baryon production cross section was obtained by computing a weighted average of the production cross sections measured for the two decay channels, $\Lc \to \proton \kam \pi^+$ and $\Lc \to \proton {\rm K^0_S} $, using the inverse of the quadratic sum of the relative statistical and uncorrelated systematic uncertainties as weights. The systematic uncertainties related to the tracking, luminosity, and generated \pt spectrum in the MC simulations are treated as correlated between the two decay channels; the uncertainty of the branching ratios is partially correlated as described in~\cite{ParticleDataGroup:2022pth}, while all the other sources of systematic uncertainties are
considered fully uncorrelated.

The production cross section integrated in \pt in the visible \pt interval of the analyses and for the results extrapolated down to $\pt=0$ are reported in Tables~\ref{tab:visible_cross_sections} and~\ref{tab:extrapolated_cross_sections}, respectively.~All the systematic uncertainties were propagated as fully correlated among the measured \pt intervals, except for the one associated with the raw-yield extraction.~The~visible production cross sections were extrapolated down to $\pt = 0 $ for non-prompt \Dzero and \Dplus mesons using FONLL predictions for beauty-hadron production in pp collisions and the PYTHIA 8 generator, employed to describe the decay kinematics of beauty hadrons ($\mathrm{h}_{\mathrm{b}}$) into charm mesons. These predictions were found to be compatible with the measurements performed in pp collisions, as shown in~\cite{ALICE:2023wbx}. In order to take into account the different system sizes with respect to pp collisions, the predictions were scaled by $A_{\mathrm{Pb}}$, assuming a flat B-meson \RpPb at unity over the whole \pt range at midrapidity, according to LHCb data~\cite{LHCb:2019avm}.
The systematic uncertainties on the extrapolation factor were estimated by considering: (i) the FONLL uncertainties, (ii) the beauty fragmentation fractions $f(\mathrm{b\to h_{b}})$, (iii) the branching ratios of the $\mathrm{h_{b} \to D + X}$ decays, and (iv) the variation of the \pt spectrum shape using EvtGen package for the description of the beauty-hadron decays~\cite{Lange:2001uf}. Contribution (ii) was estimated by considering an alternative set of beauty fragmentation fractions measured in $\mathrm{p\bar p}$ collisions~\cite{Gladilin:2014tba} while the default one is from $\mathrm{e^{+}e^{-}}$ collisions. For (iii), the branching ratios implemented in PYTHIA 8 were reweighted to reproduce the measured values reported in~\cite{ParticleDataGroup:2022pth}. It is not possible to extrapolate the non-prompt \Lc production cross section down to \pt = 0 given the absence of model calculations in \pPb collisions for beauty baryons. The prompt \Dzero has a larger contribution from resonances, like \Dstar, that do not feed the \Dplus meson. The branching ratio of B meson decays to \Dzero is significantly higher than \Dplus, where the production of $\mathrm{B^0}$ and $\mathrm{B^+}$ is similar. These factors lead to a larger production of non-prompt \Dzero mesons than the non-prompt \Dplus mesons.

\begin{table}[ht]
    \centering
        \caption{Production cross sections in the measured \pt range for non-prompt charm hadrons in \pPb collisions at \snn = 5.02~TeV.}
    \begin{tabular}{|l|cc|}
\hline
Hadron  &  Kinematic range (\GeVc)  &   $\mathrm{d}\mathit{\sigma}_{\mathrm{pPb}}^{\mathrm{visible}} / \mathrm{d}\mathit{y} |_{|\ylab|<0.5}$ ($\upmu$b)\\
\hline
\Dzero     &             $1 < \pt < 24  $&    $ 3128 \pm  183 \text{ (stat.) } \pm  187  \text{ (syst.) } \pm  116  \text{ (lumi.) }  \pm  24 \text{ (BR) }  $\\
\Dplus     &             $2 < \pt < 24  $        &    $726 \pm 101 \text{ (stat.) } \pm 42  \text{ (syst.) } \pm 27  \text{ (lumi.) }  \pm 12 \text{ (BR) }  $  \\
\Lc     &             $2 < \pt < 24  $        &  $  1404 \pm  364 \text{ (stat.) } \pm  171  \text{ (syst.) } \pm  52  \text{ (lumi.) }  \pm  70 \text{ (BR) } $   \\
\hline
    \end{tabular}
    \label{tab:visible_cross_sections}
\end{table}

\begin{table}[ht]
    \centering
       \caption{Production cross sections in the range $\pt > 0$ for non-prompt charm hadrons in \pPb collisions at \snn = 5.02~TeV.}
    \begin{tabular}{|l|cc|}
\hline
Hadron   &    Extr. factor to  &   $\mathrm{d}\sigma_{\mathrm{pPb}} / \mathrm{d}y |_{|\ylab|<0.5}$ ($\upmu$b)       \\
         &      $\pt > 0$      &                                            \\
\hline
\Dzero   &  $1.275^{+0.014}_{-0.048}$  &  $ 3990 \pm  234 \text{ (stat.) } \pm  282  \text{ (syst.) } \pm  148  \text{ (lumi.) }  \pm  30 \text{ (BR) }^{+199}_{-306} (\rm{extr.})   $  \\
\Dplus   &  $2.21^{+0.05}_{-0.19}$     &  $1604 \pm 222 \text{ (stat.) } \pm 111  \text{ (syst.) } \pm 59  \text{ (lumi.) }  \pm 27 \text{ (BR) }^{+36}_{-140} (\rm{extr.})    $  \\
\hline
    \end{tabular}
    \label{tab:extrapolated_cross_sections}
\end{table}

\subsection{Nuclear modification factors}

The nuclear modification factor $R_{\mathrm{pPb}}$ is computed as

\begin{equation}
\label{eq:nuclear_modification_factor_def}
    \RpPb = \cfrac{1}{A_{\mathrm{Pb}}} \cfrac{\d^2 \sigma_{\mathrm{pPb}} / \d \pt \d y}{\d^2 \sigma_{\mathrm{pp}} / \d \pt \d y} ~ ,
\end{equation}

where $\d^2 \sigma_{\mathrm{pPb}} / \d \pt \d y$ represents the \pt-differential production cross section within $-0.96 < \ycms < 0.04$ in \pPb collisions at $\snn = 5.02$ TeV. In the analysis of the non-prompt D mesons, $\d^2 \sigma_{\mathrm{pp}} / \d \pt \d y$ corresponds to the production cross section in pp collisions at the same center-of-mass energy at midrapidity ($|\ylab| < 0.5$), taken from~\cite{ALICE:2021mgk}. In the case of the non-prompt \Lc analysis, the pp reference was computed adopting the non-prompt \Lc $\d^2 \sigma_{\mathrm{pp}} / \d \pt \d y$ measured in pp collisions at $\s = 13$ TeV at midrapidity ($|y| < 0.5$)~\cite{ALICE:2023wbx}, scaled to account for the different collision energy. The scaling factor was computed as the ratio of the B production cross section from FONLL at $\s = 13$ and $5.02$ TeV. This is justified given: (i) the compatible \pt-dependence of the 13-to-5.02 TeV ratio of mesons and baryons in the charm sector at midrapidity~\cite{ALICE:2023sgl}, (ii) the agreement with the FONLL calculations of charm mesons at midrapidity and beauty meson at forward rapidity~\cite{ALICE:2019nxm, LHCb:2017vec}, and (iii) the assumption that the beauty baryons have a similar scaling as the beauty mesons, as observed in the charm sector~\cite{ALICE:2023wbx}. A similar behavior is found in the beauty sector~\cite{ALICE:2024xln}, though with much larger uncertainties. The shift in rapidity between pp and p–Pb collisions was corrected by using FONLL predictions for the B meson production cross sections in the two rapidity intervals, and applying the estimated correction to the non-prompt \Dzero, \Dplus, and \Lc measurements. The corresponding correction ranges from 0.9 to 2.3\%, increasing at higher \pt~\cite{ALICE:2019fhe}. The systematic uncertainties of the \pPb and \pp measurements were treated as uncorrelated within the same \pt interval and were propagated quadratically, with the exception of the BR which cancels out in the ratio.

The non-prompt \Dzero and \Dplus \RpPb are compatible over the full \pt range within the current uncertainties. In order to have a more precise measurement, the average of their \RpPb was computed, using the inverse of the quadratic sum of the statistical and uncorrelated systematic uncertainties as weights. The systematic uncertainty of the average was computed by propagating the uncertainties through the weighted average while assuming the contributions from tracking efficiency and normalisation to be fully correlated. 

\begin{figure}[ht!]
    \centering
    \begin{subfigure}{}
    \includegraphics[keepaspectratio, width=0.45\linewidth]{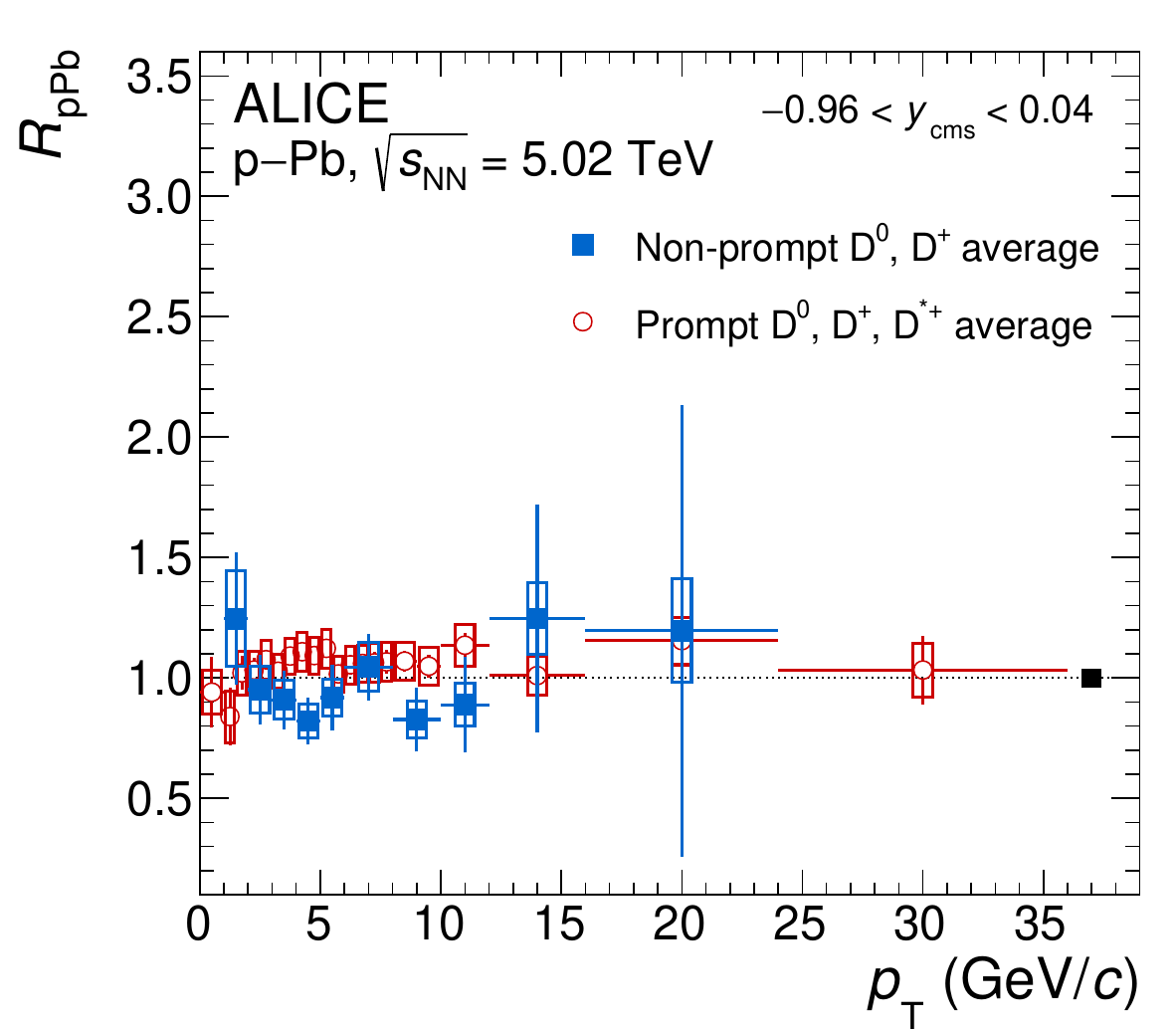}
    \end{subfigure} 
    \hfill
    \begin{subfigure}{}
    \includegraphics[keepaspectratio, width=0.45\linewidth]{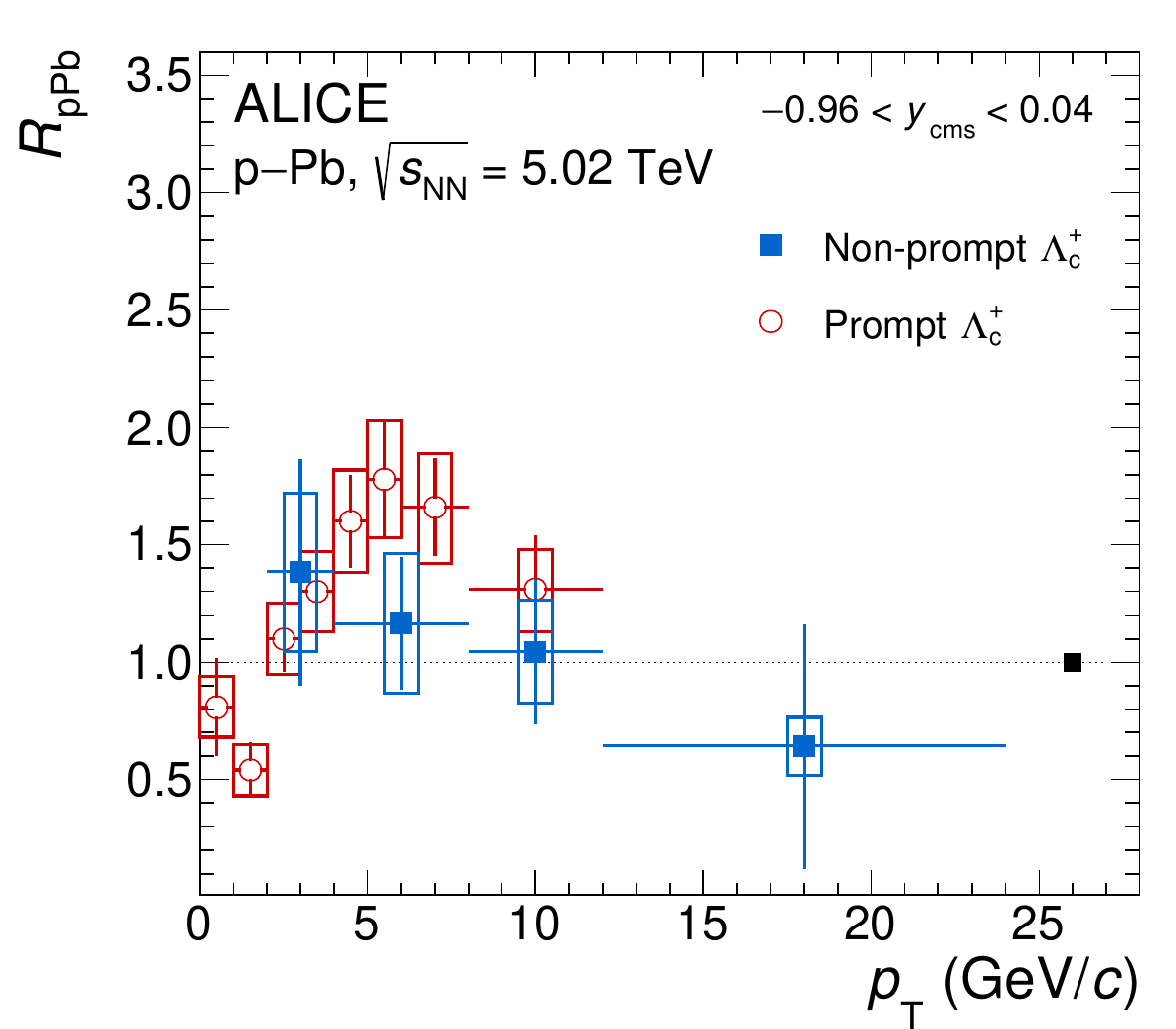}
    \end{subfigure}
    \caption{Left panel: average \pt-differential \RpPb of prompt \Dzero, \Dplus, and $\mathrm{D}^*$~\cite{ALICE:2019fhe}, and non-prompt \Dzero and \Dplus mesons.~Right panel: \pt-differential \RpPb of prompt~\cite{ALICE:2022exq} and non-prompt \Lc baryons. The vertical bars and empty boxes represent the statistical and systematic uncertainties, respectively. The black-filled box at $\RpPb = 1$ represents the normalisation systematic uncertainty.}
    \label{fig:PtDifferentialRpPbDzeroDplusLambdac}
\end{figure}

The left panel of Fig.~\ref{fig:PtDifferentialRpPbDzeroDplusLambdac} shows the average \RpPb of non-prompt \Dzero and \Dplus mesons, and the average \RpPb of prompt \Dzero, \Dplus, and \Dstar mesons~\cite{ALICE:2019fhe}.~The \RpPb of prompt and non-prompt charm mesons are compatible with each other and with unity over the entire \pt interval of the measurements within the statistical and systematic uncertainties. The comparison between the \RpPb of prompt~\cite{ALICE:2022exq} and non-prompt \Lc baryons is shown in the right panel of Fig.~\ref{fig:PtDifferentialRpPbDzeroDplusLambdac}.~The prompt \Lc-baryon \RpPb shows deviation from unity, highlighting modifications of the \pt spectrum in \pPb collisions with respect to \pp collisions, due to effects beyond nPDFs that may relate to the hadronisation process or to the presence of an expanding medium~\cite{ALICE:2020wla}. The non-prompt \Lc-baryon \RpPb is compatible both with unity and with the prompt \Lc-baryon measurements. Given its large uncertainties, it is not possible to conclude about a possible trend versus \pt.

The \pt-integrated \RpPb values for non-prompt \Dzero and \Dplus mesons in $-0.96 < \ycms < 0.04$ were calculated from the extrapolated \pt-integrated production cross sections reported above and the non-prompt \Dzero- and \Dplus-meson production cross sections measured in \pp collisions at $\sqrt{s} = 5.02$ TeV~\cite{ALICE:2021mgk}. The resulting \RpPb values of non-prompt \Dzero and \Dplus mesons in p–Pb collisions are:
\[
\begin{aligned}
    \RpPb^{\mathrm{non\text{-}prompt}\ \Dzero} (\pt > 0, -0.96 < \ycms < 0.04) &= 1.04 \pm 0.10 (\mathrm{stat.}) \pm 0.12 (\mathrm{syst.})^{+0.06}_{-0.11} (\rm{extr.}) ~ ,\\
    \RpPb^{\mathrm{non\text{-}prompt}\ \Dplus} (\pt > 0, -0.96 < \ycms < 0.04) &= 0.86 \pm 0.19 (\mathrm{stat.}) \pm 0.11 (\mathrm{syst.})^{+0.03}_{-0.11} (\rm{extr.}) ~ .
\end{aligned}
\]

The \pt-integrated \RpPb values of non-prompt D mesons are compatible with unity within uncertainties, which is consistent with a not significant modification of production cross section in \pPb collisions compared to pp collisions, as observed in the charm sector~\cite{ALICE:2023wbx}.

\begin{figure}[htb!]
    \centering
    \begin{subfigure}{}
        \includegraphics[keepaspectratio, width=0.45\linewidth]{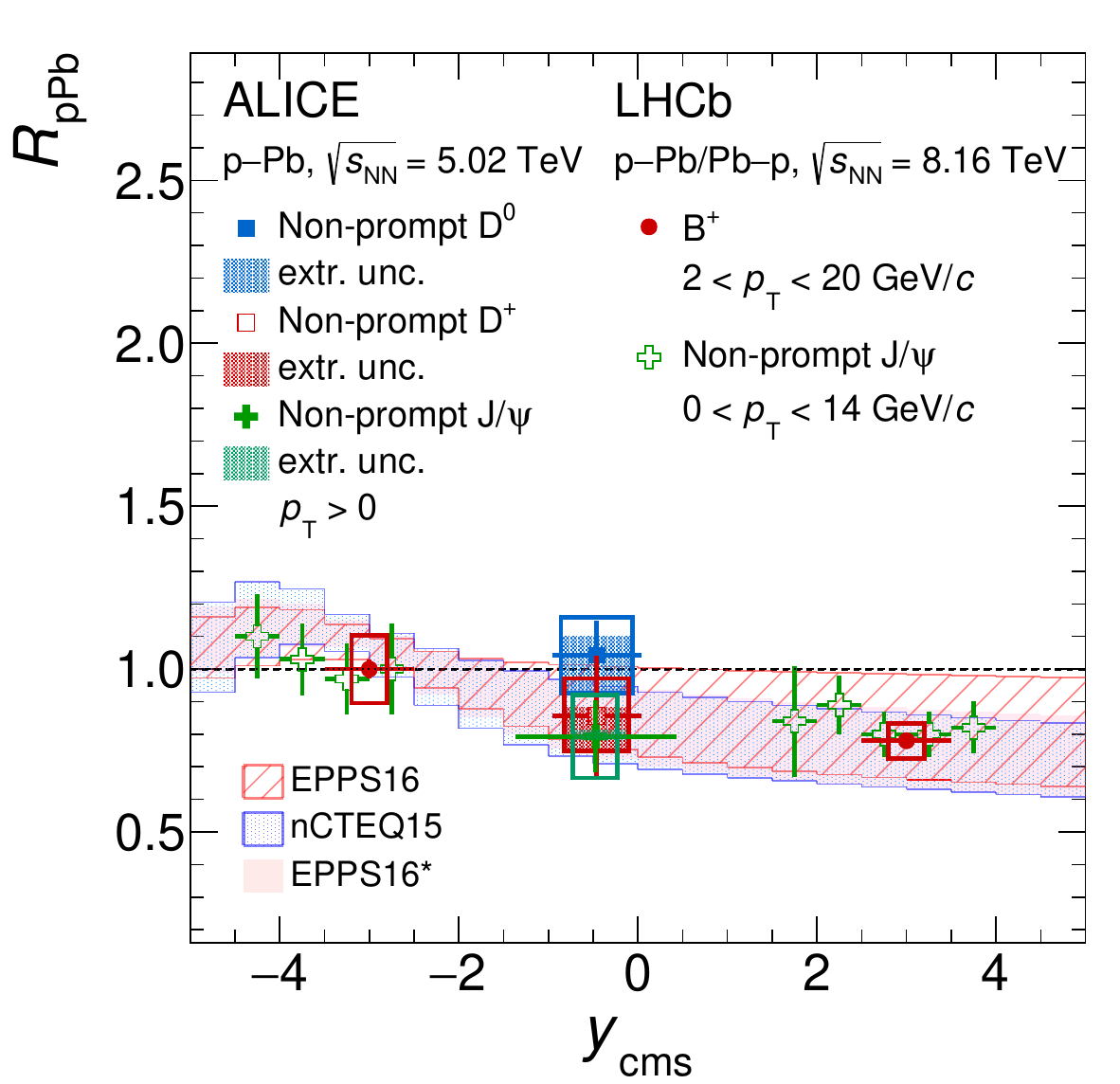}
    \end{subfigure} \hfill
    \begin{subfigure}{}
      \includegraphics[keepaspectratio, width=0.45\linewidth]{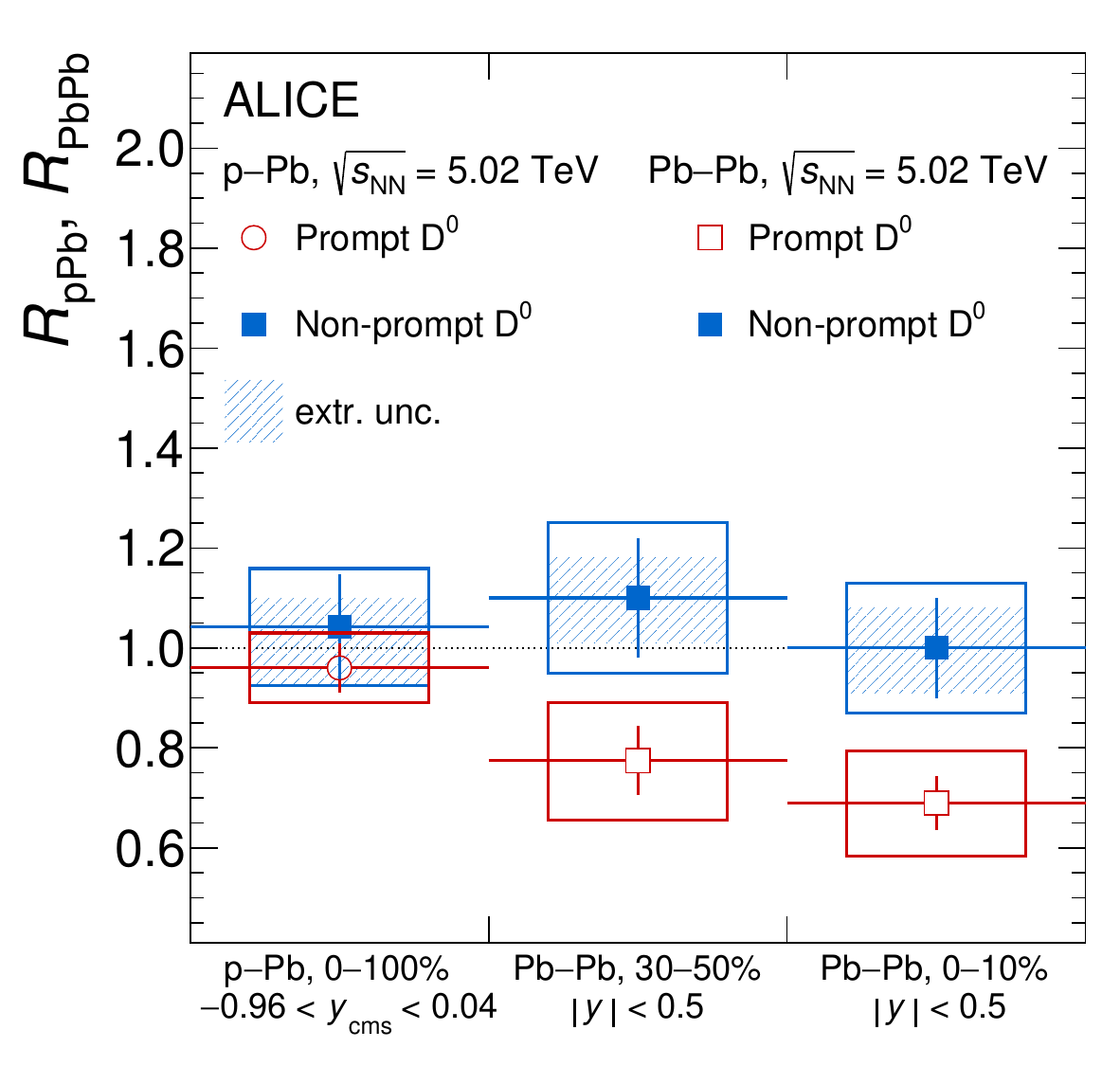}
    \end{subfigure}
    \caption{Left panel: Nuclear modification factors of non-prompt $\Dzero$ and $\Dplus$ mesons measured in p--Pb collisions at $\sqrt{s_{\rm NN}}$ = 5.02 TeV compared with the measurement of non-prompt J/$\psi$ at midrapidity~\cite{ALICE:2021lmn}, and the measurements of non-prompt J/$\psi$ and $\rm B^{+}$ mesons at forward and backward rapidity~\cite{LHCb:2019avm, LHCb:2017ygo}. The results are also compared with B-meson \RpPb calculations using different nPDF sets~\cite{Kovarik:2015cma, Eskola:2016oht, Kusina:2017gkz}.  Right panel: \pt-integrated nuclear modification factor of prompt and non-prompt $\Dzero$ mesons measured in p--Pb and Pb--Pb collisions at \snn = 5.02 TeV~\cite{ALICE:2019fhe, ALICE:2021rxa, ALICE:2022tji}. Statistical (bars) and systematic (boxes) uncertainties are shown. Extrapolation uncertainties of non-prompt \Dzero mesons in p--Pb and Pb--Pb collisions are shown separately as shaded bands.}
    \label{fig:PtIntRpPbNPModel}
\end{figure}

The left panel of Fig.~\ref{fig:PtIntRpPbNPModel} shows the \pt-integrated \RpPb measured at midrapidity for non-prompt \Dzero, \Dplus, and J/$\psi$ mesons by the ALICE Collaboration~\cite{ALICE:2021lmn}  compared to the ones of non-prompt J/$\psi$ and $\rm B^{+}$ mesons measured at forward ($1.5 < \ycms < 4.0$, $2.5 < \ycms < 3.5$) and backward ($-5.0 < \ycms < -2.5$, $-3.5 < \ycms < -2.5$) rapidity by the LHCb Collaboration~\cite{LHCb:2019avm,LHCb:2017ygo}.~The measurements of non-prompt D, J$/\psi$, and B mesons in \pPb collisions at forward, backward, and midrapidity exploring different Bjorken-$x$ regions, are sensitive to different levels of shadowing and saturation regimes. The experimental results of \pt-integrated \RpPb are compared with model calculations of the $\mathrm{B^+}$ meson in \pPb/Pb--p collisions at \snn = 8.16 TeV using the HELAC-onia generator~\cite{Shao:2012iz, Shao:2015vga, Lansberg:2016deg} with three different sets of nPDFs, i.e.~EPPS16~\cite{Eskola:2016oht}, nCTEQ15~\cite{Kovarik:2015cma}, and EPPS16*~\cite{Kusina:2017gkz}. In the calculations with EPPS16 and nCTEQ15, the model parameters are tuned to reproduce J/$\psi$ and $\psi(\text {2S})$ cross section measurements in pp collisions at the LHC~\cite{LHCb:2011zfl, CMS:2015lbl}.~A weighting based on several heavy-flavour measurements was applied on the nPDF set EPPS16~\cite{Eskola:2016oht}, to obtain the nPDF set EPPS16*, as explained in~\cite{Kusina:2017gkz}. The uncertainties in the theoretical predictions arise from those of the corresponding nPDF parameterisations. The measurements agree with the model calculations within the uncertainties.

The right panel of Fig.~\ref{fig:PtIntRpPbNPModel} shows the \pt-integrated nuclear modification factors of prompt~\cite{ALICE:2019fhe} and non-prompt \Dzero mesons measured in \pPb collisions compared with those measured in central (0-10\%)~\cite{ALICE:2021rxa} and semicentral (30-50\%)~\cite{ALICE:2022tji} \PbPb collisions.~These measurements provide an additional tool to investigate the modification of heavy-flavour production from pp to \pPb and \PbPb collisions in the beauty sector. A \pt-integrated \RpPb compatible with unity is measured for both prompt and non-prompt charm mesons, suggesting the overall CNM effects in the charm and beauty sector are similar in \pPb collisions. In \PbPb collisions, a hint of a different behaviour between charm and beauty is suggested, possibly due to a higher sensitivity of charm quarks to the nPDF modification (shadowing). Extending the measurement of beauty hadron production down to \pt= 0, both in \pPb and \PbPb collisions, will be crucial to finally achieve a complete understanding of possible modification of the heavy-flavour production due to CNM effects and possible different hadronisation mechanisms across collision systems.

\subsection{Production cross section ratios}

To probe hadronisation in \pPb collisions and its possible modification with respect to smaller collision systems, ratios of non-prompt \Dplus over non-prompt \Dzero, and non-prompt \Lc over non-prompt \Dzero \pt-integrated production cross sections were computed. The systematic uncertainties were propagated to the ratios as uncorrelated except for the ones related to tracking efficiency and normalisation, which were treated as fully correlated. The ratios are reported in Tables~\ref{tab:cross_sections_ratios} and~\ref{tab:cross_sections_ratios_2}, respectively.

The non-prompt $\Dplus/\Dzero$ \pt-integrated yield ratios are reported in Table~\ref{tab:cross_sections_ratios}, together with the values measured in pp collisions~\cite{ALICE:2021mgk} at \s = 5.02 TeV and with the one measured in \ee collisions at LEP~\cite{Gladilin:2014tba}, where the error includes the statistical uncertainties, systematic uncertainties and the uncertainties from the relevant branching fractions.
The results are compatible within experimental uncertainties, and no dependence on the collision system or energy is observed.

\begin{table}[ht]
    \centering
        \caption{Production cross section ratios of non-prompt \Dplus over \Dzero for the measured \pt ranges at midrapidty ($|\ylab| < 0.5$) in \pp collisions at \s = 5.02 TeV~\cite{ALICE:2021mgk}, \pPb collisions at \snn = 5.02 TeV, and in $\mathrm{e^{+}e^{-}}$ collisions at \s = 209 GeV at LEP~\cite{Gladilin:2014tba}.}
    \begin{tabular}{|l|cc|}
\hline
System & Kinematic range (\GeVc) & Non-prompt $\Dplus / \Dzero$   \\
\hline
\multirow{2}{*}{\pp at \s = 5.02 TeV~\cite{ALICE:2021mgk}} & \multirow{2}{*}{$2 < \pt < 24 $} & $ 0.487 \pm 0.090 \text{ (stat.)}$ \\
 &  & $ \pm \, 0.055  \text{ (syst.)} \pm 0.009 \text{ (BR)}$\\ \hline
\multirow{2}{*}{\pPb at \snn = 5.02 TeV} & \multirow{2}{*}{$2 < \pt < 24 $}  &   $ 0.402 \pm  0.060 \text{ (stat.)} $ \\ 
 &  & $\pm \, 0.034  \text{ (syst.)}  \pm 0.011 \text{ (BR)}$ \\ \hline
$\mathrm{e^{+}e^{-}}$ at \s = 209 GeV & \multirow{2}{*}{--}  & \multirow{2}{*}{$ 0.380 \pm  0.025 $} \\ LEP average~\cite{Gladilin:2014tba} & & \\
\hline
    \end{tabular}
    \label{tab:cross_sections_ratios}
\end{table}

\begin{figure}[ht!]
    \centering
    \begin{subfigure}{}
    \includegraphics[width=0.45\linewidth]{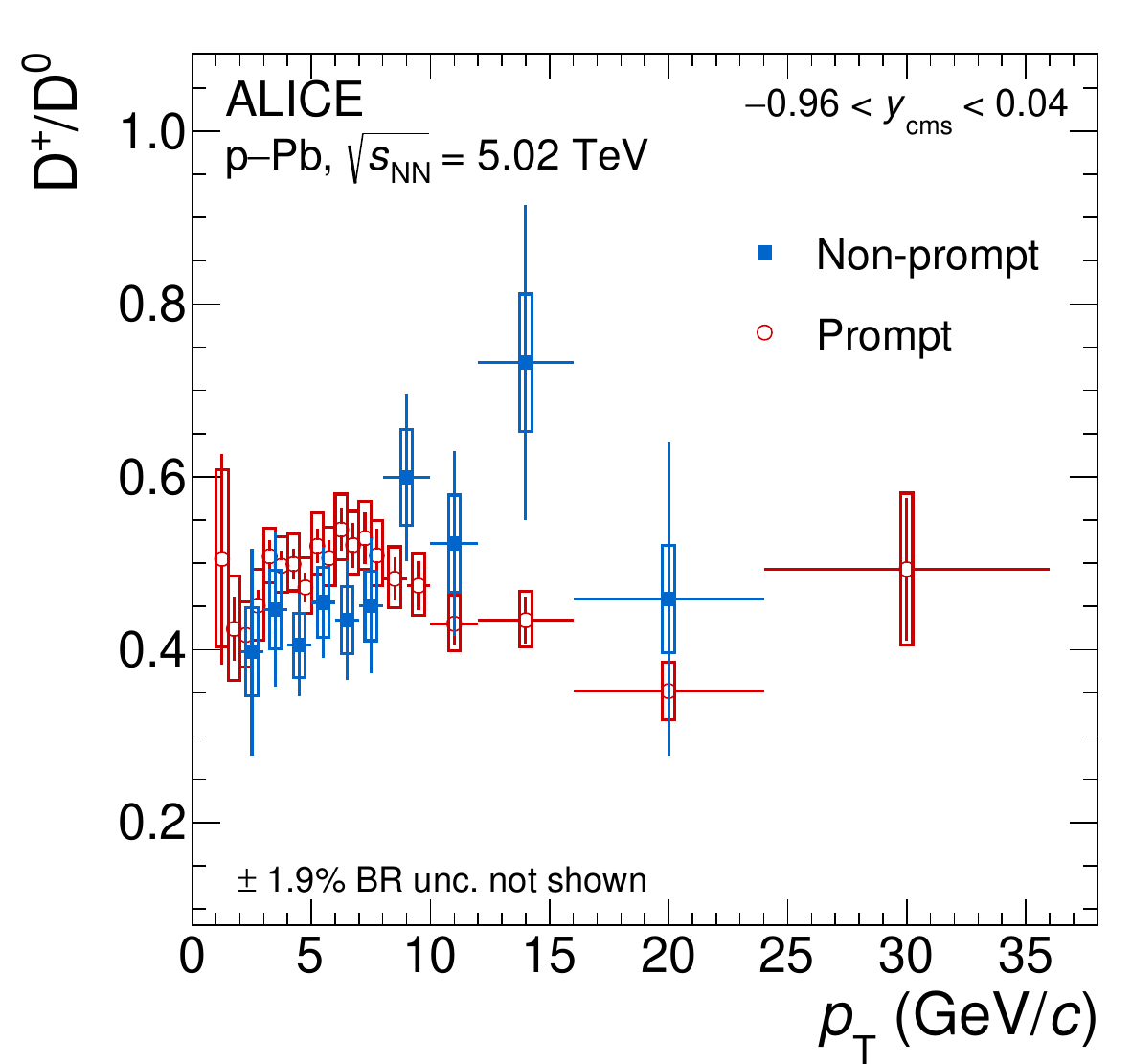}
    \end{subfigure} \hfill
    \begin{subfigure}{}
    \includegraphics[width=0.45\linewidth]{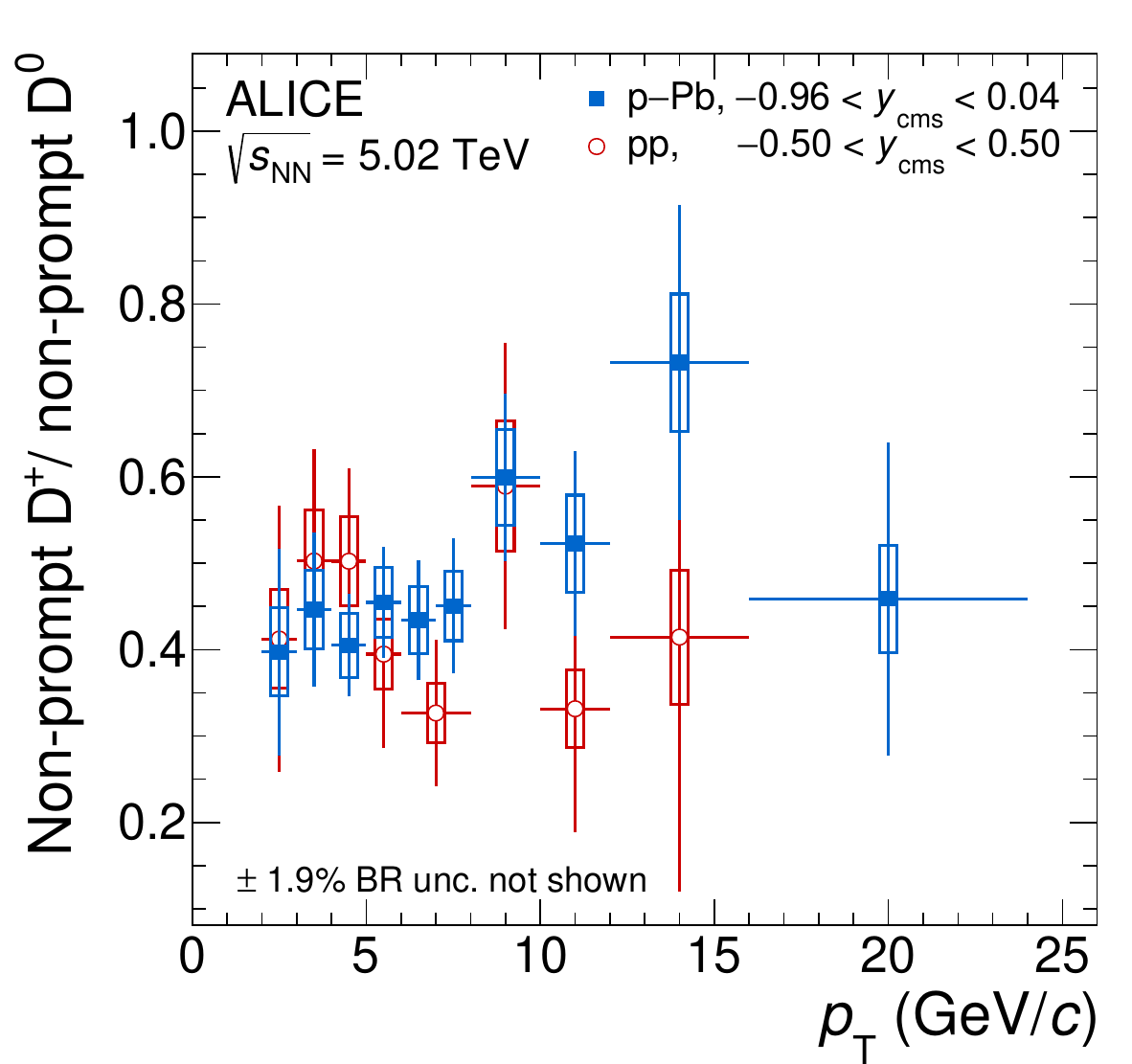}
    \end{subfigure}
    \caption{Left panel: prompt (red)~\cite{ALICE:2019fhe} and non-prompt (blue) $\Dplus/\Dzero$ yield ratio as a function of \pt. Right panel: non-prompt $\Dplus/\Dzero$ yield ratios as a function of \pt measured by ALICE Collaboration in pp (red)~\cite{ALICE:2021mgk} and p--Pb (blue) collisions at the same collision energy. The vertical bars and empty boxes represent the statistical and systematic uncertainties (without the branching ratio contribution), respectively.}
    \label{fig:CrossSectionRatios}
\end{figure}

A possible \pt dependence was investigated by computing the \pt-differential ratios.
The ratios of the \pt-differential production cross sections for prompt and non-prompt $\Dplus/\Dzero$ are shown in the left panel of Fig.~\ref{fig:CrossSectionRatios}.
The non-prompt $\Dplus/\Dzero$ ratio is independent of \pt in the measured \pt range within the current experimental precision and is compatible with the prompt $\Dplus/\Dzero$ ratio pointing to similar relative fragmentation fractions of charm and beauty quarks into D mesons. This result is in line with what was observed in the same rapidity interval in pp and \PbPb collisions at different collision energies~\cite{ALICE:2021mgk, ALICE:2021rxa}. In the right panel of Fig.~\ref{fig:CrossSectionRatios}, the non-prompt $\Dplus/\Dzero$ ratio measured in p--Pb collisions is compared with the non-prompt $\Dplus/\Dzero$ ratio measured in pp collisions at the same collision energy. The two measurements are compatible over the full \pt range of the measurements within the uncertainties, pointing to no significant modification of beauty quarks to mesons within uncertainties.

The ratio between the \pt-integrated production cross sections of non-prompt \Lc and \Dzero hadrons is reported in Table~\ref{tab:cross_sections_ratios_2}, together with the one measured in pp collisions at \s = 13 TeV ($|y| < $ 0.5)~\cite{ALICE:2023wbx} and the one measured at LEP~\cite{Gladilin:2014tba}.~Despite the different collision energies, an agreement within the experimental uncertainties is observed between the measurements performed in pp and p--Pb collisions. On the other hand, a significant difference is observed when comparing them with the \ee measurement obtained at LEP exhibiting a significant enhancement in the measured \pt range, with respect to the meson production at midrapidity in the beauty sector, as observed in the charm sector~\cite{ALICE:2023wbx}.

\begin{table}[ht]
    \centering
    \caption{Cross sections ratios of non-prompt \Lc and \Dzero for the measured \pt ranges at midrapidity ($|\ylab| < 0.5$) in \pp collisions at \s = 13 TeV~\cite{ALICE:2023wbx}, \pPb collisions at \snn = 5.02 TeV, and in $\mathrm{e^{+}e^{-}}$ collisions at \s = 209 GeV at LEP~\cite{Gladilin:2014tba}.}
    \resizebox{\textwidth}{!}{
        \begin{tabular}{|l|cc|}
            \hline
            System & Kinematic range (\GeVc)  &   Non-prompt $\Lc / \Dzero$   \\
            \hline
            \pp at \s = 13 TeV~\cite{ALICE:2023wbx}  &  $2 < \pt < 24$  &   $ 0.55 \pm 0.07 \text{ (stat.)} \pm 0.06 \text{ (syst.)}$  \\
            \hline
            \pPb at \snn = 5.02 TeV  &  $2 < \pt < 24$   &   $ 0.78 \pm 0.21 \text{ (stat.)} \pm 0.22 \text{ (syst.)}$  \\
            \hline
            $\mathrm{e^{+}e^{-}}$ at \s = 209 GeV, LEP average~\cite{Gladilin:2014tba} &  --  &   $ 0.124 \pm 0.016 $  \\
            \hline
        \end{tabular}
    }
    \label{tab:cross_sections_ratios_2}
\end{table}

\begin{figure}[ht!]
    \centering
    \begin{subfigure}{}
 \includegraphics[width=0.45\linewidth]{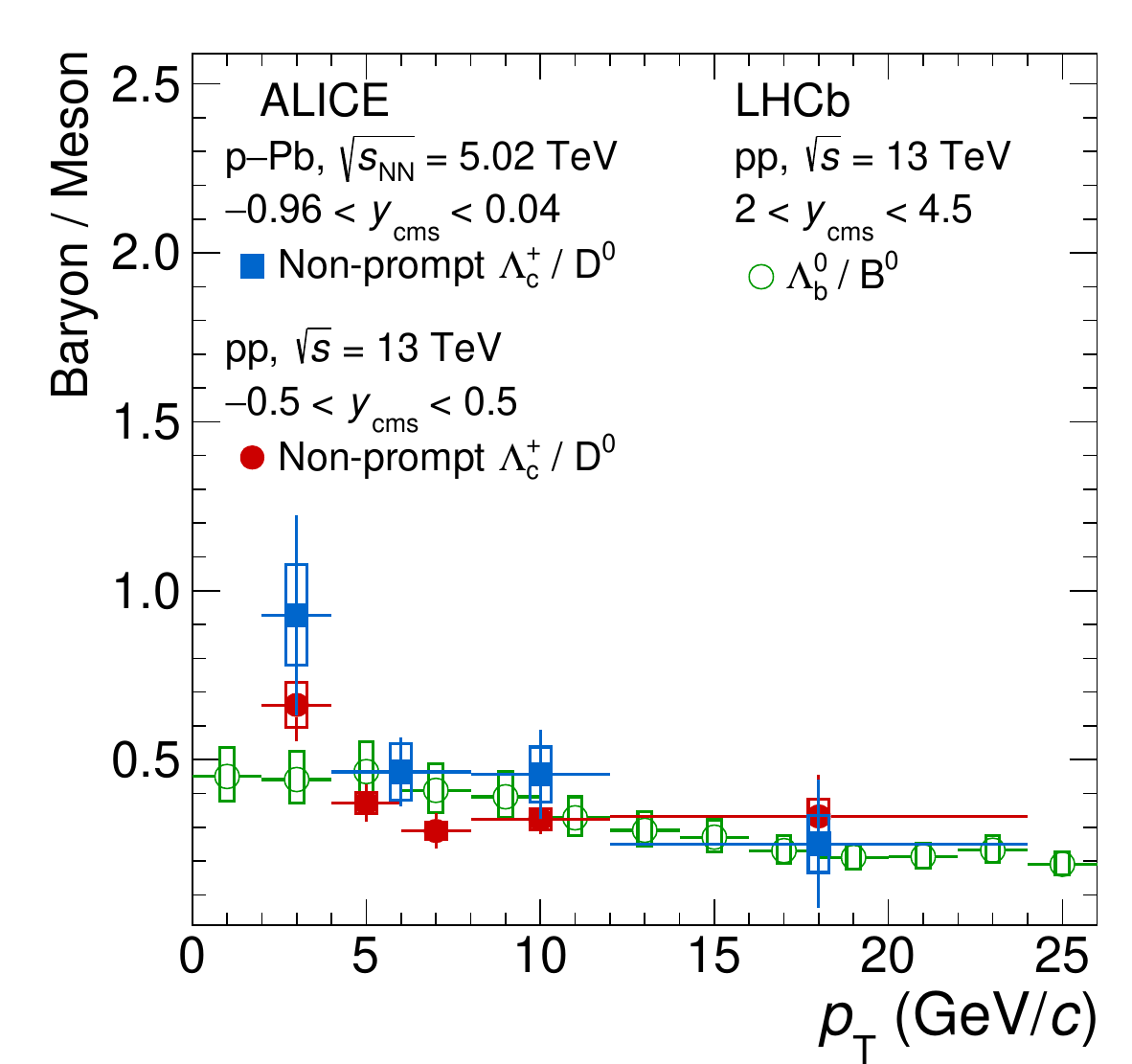}
    \end{subfigure}
    \begin{subfigure}{}
    \includegraphics[width=0.45\linewidth]{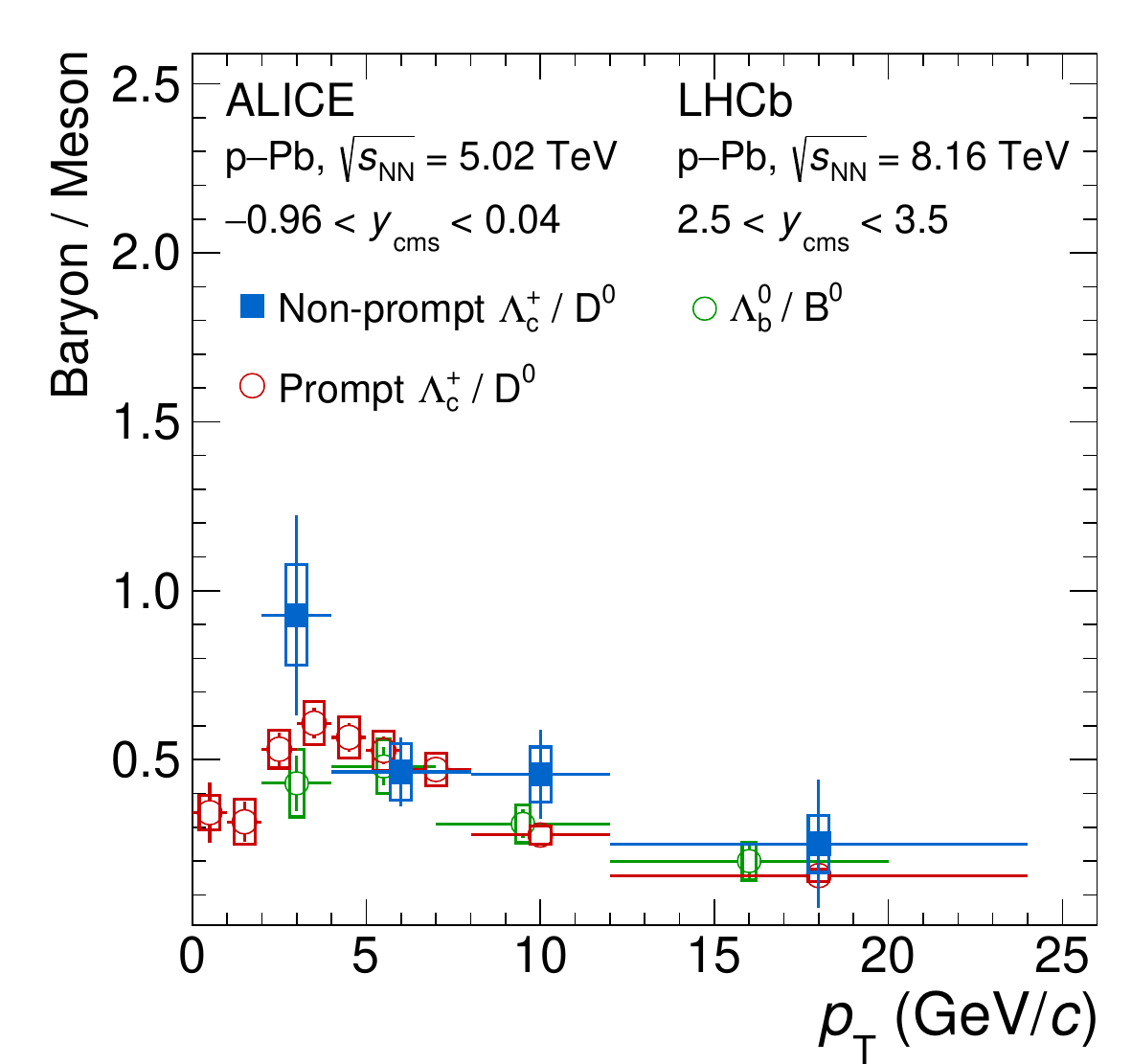}
    \end{subfigure}
    \begin{subfigure}{}
    \includegraphics[width=0.45\linewidth]{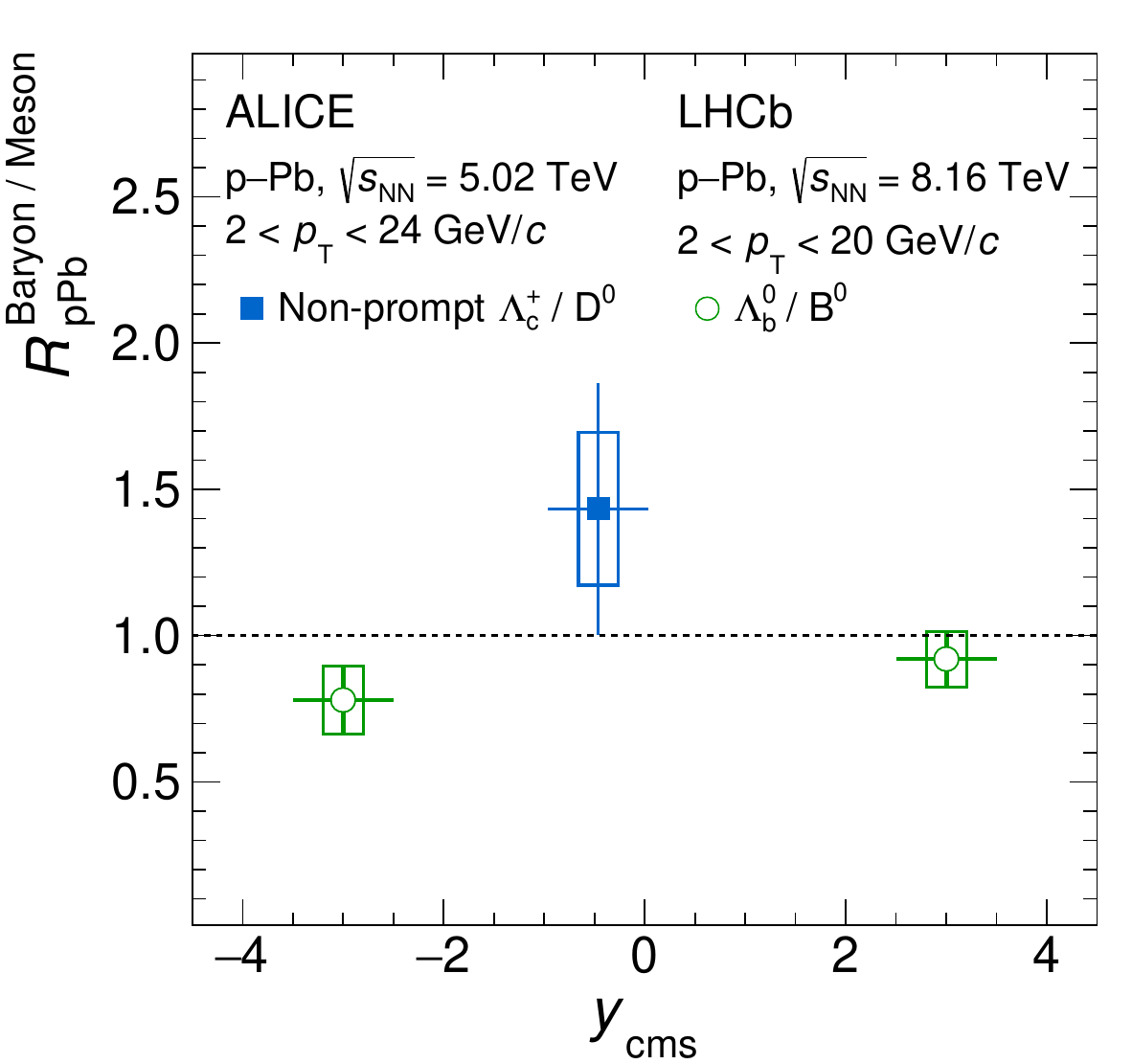}
    \end{subfigure}
    \caption{Top-left panel: non-prompt $\Lc / \Dzero$ yield ratios as a function of \pt measured by ALICE in pp (red)~\cite{ALICE:2023wbx} and p--Pb (blue) collisions compared with the \Lb$/\mathrm{B}^{0}$ ratio (green)~\cite{LHCb:2023wbo} measured by the LHCb Collaboration at forward rapidity (2 $< \ylab <$ 4.5) in pp collisions. Top-right panel: prompt (red)~\cite{ALICE:2022exq}  and non-prompt (blue) $\Lc / \Dzero$ yield ratios as a function of \pt measured by ALICE in $-0.96 < \ycms < 0.04$ together with the $\Lambda_{\mathrm{b}}^0 / \mathrm{B}^0$ yield ratio (green)~\cite{LHCb:2019avm} measured by LHCb in $2.5 < \ycms < 3.5$. Bottom panel: ratios of the nuclear modification factor of non-prompt \Lc and non-prompt \Dzero at midrapidity (blue), and the \Lb and $\mathrm{B}^{0}$ at forward and backward rapidity measured by LHCb (red)~\cite{LHCb:2019avm}.}
    \label{fig:CrossSectionRatiosLc}
\end{figure}

In order to gain further information about modification of hadronisation mechanisms in the beauty sector, the ratio of the \pt-differential production cross sections of non-prompt \Lc and \Dzero hadrons measured in p--Pb collisions at \snn = 5.02 TeV is computed and shown in the top-left panel of Fig.~\ref{fig:CrossSectionRatiosLc}. It is compared to the analogous ratio measured in pp collisions at \s = 13 TeV~\cite{ALICE:2023wbx} and with the $\Lb/\mathrm{B}^{0}$ ratio measured by the LHCb Collaboration at forward rapidity (\mbox{2 $< \ylab <$ 4.5}) in pp collisions at \s = 13 TeV~\cite{LHCb:2023wbo}. The baryon-to-meson ratio shows a decreasing trend with increasing \pt in both pp and \pPb collisions. The baryon enhancement suggested at low \pt is qualitatively similar to what was measured in pp collisions, where it was explained by different modelling of hadronisation mechanisms beyond pure in-vacuum fragmentation. Notable among these are the coalescence or recombination of charm quarks with quarks from a thermal medium~\cite{Minissale:2020bif,Song:2018tpv, Beraudo:2023nlq}, the statistical hadronisation that takes into account undiscovered higher charm resonant states~\cite{He:2019tik,He:2022tod}, and the string formation beyond the leading colour approximation~\cite{Christiansen:2015yqa,Bierlich:2023okq}.~The lack of similar models for the beauty hadrons in \pPb collisions, that could also account for the presence of the Pb nucleons in the collisions, prevents any conclusion about the origin of this modification in \pPb collisions. Neglecting a possible dependence on the collision energy, which is not observed in the charm sector~\cite{ALICE:2023sgl}, the measurement in p--Pb collisions hints at a higher non-prompt $\Lc/\Dzero$ yield ratio in 2 $< \pt <$ 12 GeV$/c$ with respect to the pp one, similarly to what is more precisely measured for the prompt $\Lc/\Dzero$~\cite{ALICE:2022exq} and $\xicz/\Dzero$ ratios~\cite{ALICE:2024ozd} and for the $\Lb/\mathrm{B}^{0}$ ratio ($-3.5  < \ylab <  -2.5$ and 2.5 $< \ylab <$ 3.5)~\cite{LHCb:2019avm}. This difference suggests a possible hardening of the beauty baryon \pt spectra, consistent with a radial flow scenario, where the shift to higher \pt depends on the particle mass. Similar spectrum modifications in \pPb collisions were also observed in the strangeness sector by the ALICE and CMS Collaborations~\cite{ALICE:2013wgn, CMS:2019isl}, and were found to be in line with the effect of radial flow predicted by hydrodynamic models such as EPOS LHC~\cite{PhysRevC.92.034906}. 
Future precise measurements down to \pt = 0 will be crucial to assess potential differences on the beauty baryon yields and collective motion in pp and \pPb collisions.

As shown in the top-right panel of Fig.~\ref{fig:CrossSectionRatiosLc}, the non-prompt baryon-to-meson ratio shows a similar trend as a function of \pt, within the uncertainties, compared to the prompt $\Lc/\Dzero$~\cite{ALICE:2022exq} in the measured \pt range in \pPb collisions at \snn = 5.02 TeV.~The results are also compared with the $\mathrm{\Lambda^{0}_{b}/B^{0}}$ yield ratio measured by the LHCb Collaboration at \mbox{$2.5 < \ycms < 3.5$}~\cite{LHCb:2019avm} in \pPb collisions at \snn = 8.16 TeV. Despite the difference in collision energy, rapidity, and the slight difference in \pt coverage between the ALICE and LHCb measurement, both sets of measurements exhibit similar dependency on \pt within experimental uncertainties. As mentioned before, the comparison between charm and beauty baryon-to-meson ratios is useful to investigate possible similarities in the hadronisation mechanism of heavy quarks. It is important to note that the additional effect of the $\mathrm{h_{b} \to h_{c} + X}$ decay kinematics is expected to slightly modify the \pt dependence of the non-prompt charm hadron ratios with respect to the beauty-hadron ratios. However, interestingly, the measurements for beauty and charm baryon-to-meson ratio show a similar trend as a function of \pt, suggesting that the hadronisation modifications for beauty quarks may mirror those for charm quarks~\cite{ALICE:2021dhb}.~The $\Lb/\mathrm{B}^{0}$ ratio from the LHCb Collaboration is lower than the non-prompt $\Lc / \Dzero$ ratio in the low-\pt interval when compared to both collision systems. However, the large experimental uncertainties over the full \pt ranges prevent from drawing strong conclusions.

Assuming that the modifications of hadronisation mechanisms of heavy quarks are similar in pp and \pPb collisions, one would expect the double ratio of non-prompt \Lc (\Lb) over \Dzero ($\mathrm{B}^{0}$) production in p--Pb to pp collisions (non-prompt $\RpPb^{\LcD}$ or $\RpPb^{\Lb/\text{B}^0}$) to be consistent with unity. This quantity corresponds to the ratio of nuclear modification factors of baryons over mesons. The bottom panel of Fig.~\ref{fig:CrossSectionRatiosLc} shows the \pt-integrated (\pt $> 2~\text{GeV/}c$) \RpPb of non-prompt \Lc baryons divided by that of non-prompt \Dzero mesons, compared with the same ratio for \Lb baryons and $\mathrm{B}^{0}$ mesons measured by the LHCb Collaboration~\cite{LHCb:2019avm} as a function of \ycms. The result in $-0.96 <\ycms < 0.04$ is consistent with unity within the uncertainties. However, more precise measurements exploiting larger collected data samples, are required to conclude on a possible rapidity dependence of beauty-baryon hadronisation.

\section{Summary}
\label{summary}

The first measurements of non-prompt \Dzero-, \Dplus-, and \Lc-hadron production at midrapidity in \pPb collisions are reported. Extrapolating the visible non-prompt D meson production cross sections down to $\pt = 0 $, the \pt-integrated \RpPb of D mesons is computed. Within the uncertainties, the \pt-integrated \RpPb of non-prompt D mesons is consistent with unity. Similarly, the \pt-integrated \RpPb of non-prompt \Dzero is compatible with prompt \Dzero. The \pt-differential \RpPb of non-prompt D mesons and \Lc is compatible with unity and the measurements in the charm sector, in the measured \pt range, within uncertainties. 
However, due to the current experimental uncertainties, it remains challenging to clearly differentiate between a flat trend compatible with unity, and the trend observed for prompt \Lc in \pPb collisions as a function of \pt.
The prompt and non-prompt yield ratios are compatible within current experimental uncertainties for both \Dplus/\Dzero and \Lc/\Dzero.
The \pt-differential non-prompt \LcD in \pPb collisions is compatible with the non-prompt \LcD ratio and \Lb/$\text{B}^0$ ratio measured in pp collisions. The \pt-integrated non-prompt $\RpPb^{\LcD}$ was measured at midrapidity and is compatible with the \pt-integrated $\RpPb^{\Lb/\text{B}^0}$ measured by the LHCb Collaboration. The results indicate no significant CNM effects in the beauty sector within uncertainties.
These novel measurements in \pPb collisions provide important insights, enriching the understanding of nPDF models and the modification of beauty quark hadronisation mechanisms.~In addition, these measurements represent an important input for constraining theoretical models for heavy-flavour hadron production in \pPb collisions, which are still lacking at the moment. With the major upgrade of the ALICE detector for Run 3, larger data samples, and foreseen upgrades for Run 4, ALICE will significantly advance this field in the near future.

\newenvironment{acknowledgement}{\relax}{\relax}
\begin{acknowledgement}
\section*{Acknowledgements}

The ALICE Collaboration would like to thank all its engineers and technicians for their invaluable contributions to the construction of the experiment and the CERN accelerator teams for the outstanding performance of the LHC complex.
The ALICE Collaboration gratefully acknowledges the resources and support provided by all Grid centres and the Worldwide LHC Computing Grid (WLCG) collaboration.
The ALICE Collaboration acknowledges the following funding agencies for their support in building and running the ALICE detector:
A. I. Alikhanyan National Science Laboratory (Yerevan Physics Institute) Foundation (ANSL), State Committee of Science and World Federation of Scientists (WFS), Armenia;
Austrian Academy of Sciences, Austrian Science Fund (FWF): [M 2467-N36] and Nationalstiftung f\"{u}r Forschung, Technologie und Entwicklung, Austria;
Ministry of Communications and High Technologies, National Nuclear Research Center, Azerbaijan;
Conselho Nacional de Desenvolvimento Cient\'{\i}fico e Tecnol\'{o}gico (CNPq), Financiadora de Estudos e Projetos (Finep), Funda\c{c}\~{a}o de Amparo \`{a} Pesquisa do Estado de S\~{a}o Paulo (FAPESP) and Universidade Federal do Rio Grande do Sul (UFRGS), Brazil;
Bulgarian Ministry of Education and Science, within the National Roadmap for Research Infrastructures 2020-2027 (object CERN), Bulgaria;
Ministry of Education of China (MOEC) , Ministry of Science \& Technology of China (MSTC) and National Natural Science Foundation of China (NSFC), China;
Ministry of Science and Education and Croatian Science Foundation, Croatia;
Centro de Aplicaciones Tecnol\'{o}gicas y Desarrollo Nuclear (CEADEN), Cubaenerg\'{\i}a, Cuba;
Ministry of Education, Youth and Sports of the Czech Republic, Czech Republic;
The Danish Council for Independent Research | Natural Sciences, the VILLUM FONDEN and Danish National Research Foundation (DNRF), Denmark;
Helsinki Institute of Physics (HIP), Finland;
Commissariat \`{a} l'Energie Atomique (CEA) and Institut National de Physique Nucl\'{e}aire et de Physique des Particules (IN2P3) and Centre National de la Recherche Scientifique (CNRS), France;
Bundesministerium f\"{u}r Bildung und Forschung (BMBF) and GSI Helmholtzzentrum f\"{u}r Schwerionenforschung GmbH, Germany;
General Secretariat for Research and Technology, Ministry of Education, Research and Religions, Greece;
National Research, Development and Innovation Office, Hungary;
Department of Atomic Energy Government of India (DAE), Department of Science and Technology, Government of India (DST), University Grants Commission, Government of India (UGC) and Council of Scientific and Industrial Research (CSIR), India;
National Research and Innovation Agency - BRIN, Indonesia;
Istituto Nazionale di Fisica Nucleare (INFN), Italy;
Japanese Ministry of Education, Culture, Sports, Science and Technology (MEXT) and Japan Society for the Promotion of Science (JSPS) KAKENHI, Japan;
Consejo Nacional de Ciencia (CONACYT) y Tecnolog\'{i}a, through Fondo de Cooperaci\'{o}n Internacional en Ciencia y Tecnolog\'{i}a (FONCICYT) and Direcci\'{o}n General de Asuntos del Personal Academico (DGAPA), Mexico;
Nederlandse Organisatie voor Wetenschappelijk Onderzoek (NWO), Netherlands;
The Research Council of Norway, Norway;
Pontificia Universidad Cat\'{o}lica del Per\'{u}, Peru;
Ministry of Science and Higher Education, National Science Centre and WUT ID-UB, Poland;
Korea Institute of Science and Technology Information and National Research Foundation of Korea (NRF), Republic of Korea;
Ministry of Education and Scientific Research, Institute of Atomic Physics, Ministry of Research and Innovation and Institute of Atomic Physics and Universitatea Nationala de Stiinta si Tehnologie Politehnica Bucuresti, Romania;
Ministry of Education, Science, Research and Sport of the Slovak Republic, Slovakia;
National Research Foundation of South Africa, South Africa;
Swedish Research Council (VR) and Knut \& Alice Wallenberg Foundation (KAW), Sweden;
European Organization for Nuclear Research, Switzerland;
Suranaree University of Technology (SUT), National Science and Technology Development Agency (NSTDA) and National Science, Research and Innovation Fund (NSRF via PMU-B B05F650021), Thailand;
Turkish Energy, Nuclear and Mineral Research Agency (TENMAK), Turkey;
National Academy of  Sciences of Ukraine, Ukraine;
Science and Technology Facilities Council (STFC), United Kingdom;
National Science Foundation of the United States of America (NSF) and United States Department of Energy, Office of Nuclear Physics (DOE NP), United States of America.
In addition, individual groups or members have received support from:
Czech Science Foundation (grant no. 23-07499S), Czech Republic;
European Research Council (grant no. 950692), European Union;
ICSC - Centro Nazionale di Ricerca in High Performance Computing, Big Data and Quantum Computing, European Union - NextGenerationEU;
Academy of Finland (Center of Excellence in Quark Matter) (grant nos. 346327, 346328), Finland.

\end{acknowledgement}

\bibliographystyle{utphys}
\bibliography{bibliography}

\newpage
\appendix

\section{The ALICE Collaboration}
\label{app:collab}
\begin{flushleft} 
\small

S.~Acharya\,\orcidlink{0000-0002-9213-5329}\,$^{\rm 127}$, 
D.~Adamov\'{a}\,\orcidlink{0000-0002-0504-7428}\,$^{\rm 86}$, 
A.~Agarwal$^{\rm 135}$, 
G.~Aglieri Rinella\,\orcidlink{0000-0002-9611-3696}\,$^{\rm 32}$, 
L.~Aglietta\,\orcidlink{0009-0003-0763-6802}\,$^{\rm 24}$, 
M.~Agnello\,\orcidlink{0000-0002-0760-5075}\,$^{\rm 29}$, 
N.~Agrawal\,\orcidlink{0000-0003-0348-9836}\,$^{\rm 25}$, 
Z.~Ahammed\,\orcidlink{0000-0001-5241-7412}\,$^{\rm 135}$, 
S.~Ahmad\,\orcidlink{0000-0003-0497-5705}\,$^{\rm 15}$, 
S.U.~Ahn\,\orcidlink{0000-0001-8847-489X}\,$^{\rm 71}$
% ...

\section*{Affiliation Notes}

$^{\rm I}$ Deceased\\
$^{\rm II}$ Also at: Max-Planck-Institut fur Physik, Munich, Germany\\
$^{\rm III}$ Also at: Italian National Agency for New Technologies, Energy and Sustainable Economic Development (ENEA), Bologna, Italy\\
$^{\rm IV}$ Also at: Dipartimento DET del Politecnico di Torino, Turin, Italy\\
$^{\rm V}$ Also at: Yildiz Technical University, Istanbul, T\"{u}rkiye\\
$^{\rm VI}$ Also at: Department of Applied Physics, Aligarh Muslim University, Aligarh, India\\
$^{\rm VII}$ Also at: Institute of Theoretical Physics, University of Wroclaw, Poland\\
$^{\rm VIII}$ Also at: An institution covered by a cooperation agreement with CERN\\

\section*{Collaboration Institutes}

$^{1}$ A.I. Alikhanyan National Science Laboratory (Yerevan Physics Institute) Foundation, Yerevan, Armenia\\
$^{2}$ AGH University of Krakow, Cracow, Poland\\
$^{3}$ Bogolyubov Institute for Theoretical Physics, National Academy of Sciences of Ukraine, Kiev, Ukraine\\
$^{4}$ Bose Institute, Department of Physics  and Centre for Astroparticle Physics and Space Science (CAPSS), Kolkata, India\\
$^{5}$ California Polytechnic State University, San Luis Obispo, California, United States\\
$^{6}$ Central China Normal University, Wuhan, China\\
$^{7}$ Centro de Aplicaciones Tecnol\'{o}gicas y Desarrollo Nuclear (CEADEN), Havana, Cuba\\
$^{8}$ Centro de Investigaci\'{o}n y de Estudios Avanzados (CINVESTAV), Mexico City and M\'{e}rida, Mexico\\
$^{9}$ Chicago State University, Chicago, Illinois, United States\\
$^{10}$ China Institute of Atomic Energy, Beijing, China\\
$^{11}$ China University of Geosciences, Wuhan, China\\
$^{12}$ Chungbuk National University, Cheongju, Republic of Korea\\
$^{13}$ Comenius University Bratislava, Faculty of Mathematics, Physics and Informatics, Bratislava, Slovak Republic\\
$^{14}$ Creighton University, Omaha, Nebraska, United States\\
$^{15}$ Department of Physics, Aligarh Muslim University, Aligarh, India\\
$^{16}$ Department of Physics, Pusan National University, Pusan, Republic of Korea\\
$^{17}$ Department of Physics, Sejong University, Seoul, Republic of Korea\\
$^{18}$ Department of Physics, University of California, Berkeley, California, United States\\
$^{19}$ Department of Physics, University of Oslo, Oslo, Norway\\
$^{20}$ Department of Physics and Technology, University of Bergen, Bergen, Norway\\
$^{21}$ Dipartimento di Fisica, Universit\`{a} di Pavia, Pavia, Italy\\
$^{22}$ Dipartimento di Fisica dell'Universit\`{a} and Sezione INFN, Cagliari, Italy\\
$^{23}$ Dipartimento di Fisica dell'Universit\`{a} and Sezione INFN, Trieste, Italy\\
$^{24}$ Dipartimento di Fisica dell'Universit\`{a} and Sezione INFN, Turin, Italy\\
$^{25}$ Dipartimento di Fisica e Astronomia dell'Universit\`{a} and Sezione INFN, Bologna, Italy\\
$^{26}$ Dipartimento di Fisica e Astronomia dell'Universit\`{a} and Sezione INFN, Catania, Italy\\
$^{27}$ Dipartimento di Fisica e Astronomia dell'Universit\`{a} and Sezione INFN, Padova, Italy\\
$^{28}$ Dipartimento di Fisica `E.R.~Caianiello' dell'Universit\`{a} and Gruppo Collegato INFN, Salerno, Italy\\
$^{29}$ Dipartimento DISAT del Politecnico and Sezione INFN, Turin, Italy\\
$^{30}$ Dipartimento di Scienze MIFT, Universit\`{a} di Messina, Messina, Italy\\
$^{31}$ Dipartimento Interateneo di Fisica `M.~Merlin' and Sezione INFN, Bari, Italy\\
$^{32}$ European Organization for Nuclear Research (CERN), Geneva, Switzerland\\
$^{33}$ Faculty of Electrical Engineering, Mechanical Engineering and Naval Architecture, University of Split, Split, Croatia\\
$^{34}$ Faculty of Engineering and Science, Western Norway University of Applied Sciences, Bergen, Norway\\
$^{35}$ Faculty of Nuclear Sciences and Physical Engineering, Czech Technical University in Prague, Prague, Czech Republic\\
$^{36}$ Faculty of Physics, Sofia University, Sofia, Bulgaria\\
$^{37}$ Faculty of Science, P.J.~\v{S}af\'{a}rik University, Ko\v{s}ice, Slovak Republic\\
$^{38}$ Frankfurt Institute for Advanced Studies, Johann Wolfgang Goethe-Universit\"{a}t Frankfurt, Frankfurt, Germany\\
$^{39}$ Fudan University, Shanghai, China\\
$^{40}$ Gangneung-Wonju National University, Gangneung, Republic of Korea\\
$^{41}$ Gauhati University, Department of Physics, Guwahati, India\\
$^{42}$ Helmholtz-Institut f\"{u}r Strahlen- und Kernphysik, Rheinische Friedrich-Wilhelms-Universit\"{a}t Bonn, Bonn, Germany\\
$^{43}$ Helsinki Institute of Physics (HIP), Helsinki, Finland\\
$^{44}$ High Energy Physics Group,  Universidad Aut\'{o}noma de Puebla, Puebla, Mexico\\
$^{45}$ Horia Hulubei National Institute of Physics and Nuclear Engineering, Bucharest, Romania\\
$^{46}$ HUN-REN Wigner Research Centre for Physics, Budapest, Hungary\\
$^{47}$ Indian Institute of Technology Bombay (IIT), Mumbai, India\\
$^{48}$ Indian Institute of Technology Indore, Indore, India\\
$^{49}$ INFN, Laboratori Nazionali di Frascati, Frascati, Italy\\
$^{50}$ INFN, Sezione di Bari, Bari, Italy\\
$^{51}$ INFN, Sezione di Bologna, Bologna, Italy\\
$^{52}$ INFN, Sezione di Cagliari, Cagliari, Italy\\
$^{53}$ INFN, Sezione di Catania, Catania, Italy\\
$^{54}$ INFN, Sezione di Padova, Padova, Italy\\
$^{55}$ INFN, Sezione di Pavia, Pavia, Italy\\
$^{56}$ INFN, Sezione di Torino, Turin, Italy\\
$^{57}$ INFN, Sezione di Trieste, Trieste, Italy\\
$^{58}$ Inha University, Incheon, Republic of Korea\\
$^{59}$ Institute for Gravitational and Subatomic Physics (GRASP), Utrecht University/Nikhef, Utrecht, Netherlands\\
$^{60}$ Institute of Experimental Physics, Slovak Academy of Sciences, Ko\v{s}ice, Slovak Republic\\
$^{61}$ Institute of Physics, Homi Bhabha National Institute, Bhubaneswar, India\\
$^{62}$ Institute of Physics of the Czech Academy of Sciences, Prague, Czech Republic\\
$^{63}$ Institute of Space Science (ISS), Bucharest, Romania\\
$^{64}$ Institut f\"{u}r Kernphysik, Johann Wolfgang Goethe-Universit\"{a}t Frankfurt, Frankfurt, Germany\\
$^{65}$ Instituto de Ciencias Nucleares, Universidad Nacional Aut\'{o}noma de M\'{e}xico, Mexico City, Mexico\\
$^{66}$ Instituto de F\'{i}sica, Universidade Federal do Rio Grande do Sul (UFRGS), Porto Alegre, Brazil\\
$^{67}$ Instituto de F\'{\i}sica, Universidad Nacional Aut\'{o}noma de M\'{e}xico, Mexico City, Mexico\\
$^{68}$ iThemba LABS, National Research Foundation, Somerset West, South Africa\\
$^{69}$ Jeonbuk National University, Jeonju, Republic of Korea\\
$^{70}$ Johann-Wolfgang-Goethe Universit\"{a}t Frankfurt Institut f\"{u}r Informatik, Fachbereich Informatik und Mathematik, Frankfurt, Germany\\
$^{71}$ Korea Institute of Science and Technology Information, Daejeon, Republic of Korea\\
$^{72}$ KTO Karatay University, Konya, Turkey\\
$^{73}$ Laboratoire de Physique Subatomique et de Cosmologie, Universit\'{e} Grenoble-Alpes, CNRS-IN2P3, Grenoble, France\\
$^{74}$ Lawrence Berkeley National Laboratory, Berkeley, California, United States\\
$^{75}$ Lund University Department of Physics, Division of Particle Physics, Lund, Sweden\\
$^{76}$ Nagasaki Institute of Applied Science, Nagasaki, Japan\\
$^{77}$ Nara Women{'}s University (NWU), Nara, Japan\\
$^{78}$ National and Kapodistrian University of Athens, School of Science, Department of Physics , Athens, Greece\\
$^{79}$ National Centre for Nuclear Research, Warsaw, Poland\\
$^{80}$ National Institute of Science Education and Research, Homi Bhabha National Institute, Jatni, India\\
$^{81}$ National Nuclear Research Center, Baku, Azerbaijan\\
$^{82}$ National Research and Innovation Agency - BRIN, Jakarta, Indonesia\\
$^{83}$ Niels Bohr Institute, University of Copenhagen, Copenhagen, Denmark\\
$^{84}$ Nikhef, National institute for subatomic physics, Amsterdam, Netherlands\\
$^{85}$ Nuclear Physics Group, STFC Daresbury Laboratory, Daresbury, United Kingdom\\
$^{86}$ Nuclear Physics Institute of the Czech Academy of Sciences, Husinec-\v{R}e\v{z}, Czech Republic\\
$^{87}$ Oak Ridge National Laboratory, Oak Ridge, Tennessee, United States\\
$^{88}$ Ohio State University, Columbus, Ohio, United States\\
$^{89}$ Physics department, Faculty of science, University of Zagreb, Zagreb, Croatia\\
$^{90}$ Physics Department, Panjab University, Chandigarh, India\\
$^{91}$ Physics Department, University of Jammu, Jammu, India\\
$^{92}$ Physics Program and International Institute for Sustainability with Knotted Chiral Meta Matter (SKCM2), Hiroshima University, Hiroshima, Japan\\
$^{93}$ Physikalisches Institut, Eberhard-Karls-Universit\"{a}t T\"{u}bingen, T\"{u}bingen, Germany\\
$^{94}$ Physikalisches Institut, Ruprecht-Karls-Universit\"{a}t Heidelberg, Heidelberg, Germany\\
$^{95}$ Physik Department, Technische Universit\"{a}t M\"{u}nchen, Munich, Germany\\
$^{96}$ Politecnico di Bari and Sezione INFN, Bari, Italy\\
$^{97}$ Research Division and ExtreMe Matter Institute EMMI, GSI Helmholtzzentrum f\"ur Schwerionenforschung GmbH, Darmstadt, Germany\\
$^{98}$ Saga University, Saga, Japan\\
$^{99}$ Saha Institute of Nuclear Physics, Homi Bhabha National Institute, Kolkata, India\\
$^{100}$ School of Physics and Astronomy, University of Birmingham, Birmingham, United Kingdom\\
$^{101}$ Secci\'{o}n F\'{\i}sica, Departamento de Ciencias, Pontificia Universidad Cat\'{o}lica del Per\'{u}, Lima, Peru\\
$^{102}$ Stefan Meyer Institut f\"{u}r Subatomare Physik (SMI), Vienna, Austria\\
$^{103}$ SUBATECH, IMT Atlantique, Nantes Universit\'{e}, CNRS-IN2P3, Nantes, France\\
$^{104}$ Sungkyunkwan University, Suwon City, Republic of Korea\\
$^{105}$ Suranaree University of Technology, Nakhon Ratchasima, Thailand\\
$^{106}$ Technical University of Ko\v{s}ice, Ko\v{s}ice, Slovak Republic\\
$^{107}$ The Henryk Niewodniczanski Institute of Nuclear Physics, Polish Academy of Sciences, Cracow, Poland\\
$^{108}$ The University of Texas at Austin, Austin, Texas, United States\\
$^{109}$ Universidad Aut\'{o}noma de Sinaloa, Culiac\'{a}n, Mexico\\
$^{110}$ Universidade de S\~{a}o Paulo (USP), S\~{a}o Paulo, Brazil\\
$^{111}$ Universidade Estadual de Campinas (UNICAMP), Campinas, Brazil\\
$^{112}$ Universidade Federal do ABC, Santo Andre, Brazil\\
$^{113}$ Universitatea Nationala de Stiinta si Tehnologie Politehnica Bucuresti, Bucharest, Romania\\
$^{114}$ University of Cape Town, Cape Town, South Africa\\
$^{115}$ University of Derby, Derby, United Kingdom\\
$^{116}$ University of Houston, Houston, Texas, United States\\
$^{117}$ University of Jyv\"{a}skyl\"{a}, Jyv\"{a}skyl\"{a}, Finland\\
$^{118}$ University of Kansas, Lawrence, Kansas, United States\\
$^{119}$ University of Liverpool, Liverpool, United Kingdom\\
$^{120}$ University of Science and Technology of China, Hefei, China\\
$^{121}$ University of South-Eastern Norway, Kongsberg, Norway\\
$^{122}$ University of Tennessee, Knoxville, Tennessee, United States\\
$^{123}$ University of the Witwatersrand, Johannesburg, South Africa\\
$^{124}$ University of Tokyo, Tokyo, Japan\\
$^{125}$ University of Tsukuba, Tsukuba, Japan\\
$^{126}$ Universit\"{a}t M\"{u}nster, Institut f\"{u}r Kernphysik, M\"{u}nster, Germany\\
$^{127}$ Universit\'{e} Clermont Auvergne, CNRS/IN2P3, LPC, Clermont-Ferrand, France\\
$^{128}$ Universit\'{e} de Lyon, CNRS/IN2P3, Institut de Physique des 2 Infinis de Lyon, Lyon, France\\
$^{129}$ Universit\'{e} de Strasbourg, CNRS, IPHC UMR 7178, F-67000 Strasbourg, France, Strasbourg, France\\
$^{130}$ Universit\'{e} Paris-Saclay, Centre d'Etudes de Saclay (CEA), IRFU, D\'{e}partment de Physique Nucl\'{e}aire (DPhN), Saclay, France\\
$^{131}$ Universit\'{e}  Paris-Saclay, CNRS/IN2P3, IJCLab, Orsay, France\\
$^{132}$ Universit\`{a} degli Studi di Foggia, Foggia, Italy\\
$^{133}$ Universit\`{a} del Piemonte Orientale, Vercelli, Italy\\
$^{134}$ Universit\`{a} di Brescia, Brescia, Italy\\
$^{135}$ Variable Energy Cyclotron Centre, Homi Bhabha National Institute, Kolkata, India\\
$^{136}$ Warsaw University of Technology, Warsaw, Poland\\
$^{137}$ Wayne State University, Detroit, Michigan, United States\\
$^{138}$ Yale University, New Haven, Connecticut, United States\\
$^{139}$ Yonsei University, Seoul, Republic of Korea\\
$^{140}$  Zentrum  f\"{u}r Technologie und Transfer (ZTT), Worms, Germany\\
$^{141}$ Affiliated with an institute covered by a cooperation agreement with CERN\\
$^{142}$ Affiliated with an international laboratory covered by a cooperation agreement with CERN.\\

\end{flushleft} 

\end{document}